\documentclass[12pt,preprint]{aastex}

\begin{document}

\title{The Shape and Profile of the Milky Way Halo as Seen by the Canada-France-Hawaii Telescope Legacy Survey}

\author{
Branimir Sesar\altaffilmark{\ref{Washington},\ref{Caltech}},
Mario Juri\'{c}\altaffilmark{\ref{CfA},\ref{Hubble}} \&
\v{Z}eljko Ivezi\'{c}\altaffilmark{\ref{Washington},\ref{UZg}}
}

\altaffiltext{1}{Department of Astronomy, University of Washington, P.O.~Box
                 351580, Seattle, WA 98195-1580, USA\label{Washington}}
\altaffiltext{2}{Division of Physics, Mathematics and Astronomy, California
                 Institute of Technology, Pasadena, CA 91125, USA; bsesar@astro.caltech.edu\label{Caltech}}
\altaffiltext{3}{Harvard-Smithsonian Center for Astrophysics, MS-72,
                 60 Garden Street, Cambridge, MA 02138, USA\label{CfA}}
\altaffiltext{4}{Hubble Fellow\label{Hubble}}
\altaffiltext{5}{Department of Physics, University of Zagreb, Bijeni\v{c}ka
c.~32, P.P. 331, Zagreb, Croatia\label{UZg}}

\begin{abstract}
We use Canada$-$France$-$Hawaii Telescope Legacy Survey data for 170 deg$^2$,
recalibrated and transformed to the Sloan Digital Sky Survey $ugri$ photometric
system, to study the distribution of near-turnoff main-sequence stars in the
Galactic halo along four lines of sight to heliocentric distances of $\sim35$
kpc. We find that the halo stellar number density profile becomes steeper at
Galactocentric distances greater than $R_{\rm gal}\sim28$ kpc, with the power
law index changing from $n_{\rm inner}=-2.62\pm0.04$ to
$n_{\rm outer}=-3.8\pm0.1$. In particular, we test a series of single power law
models and find them to be strongly disfavored by the data. The parameters for
the best-fit Einasto profile are $n=2.2\pm0.2$ and $R_{\rm e}=22.2\pm0.4$ kpc.
We measure the oblateness of the halo to be $q\equiv c/a = 0.70\pm0.01$ and
detect no evidence of it changing across the range of probed distances. The
Sagittarius stream is detected in the $l=173\arcdeg$ and $b=-62\arcdeg$
direction as an overdensity of ${\rm [Fe/H]}\sim-1.5$ dex stars at
$R_{\rm gal}\sim32$ kpc, providing a new constraint for the Sagittarius stream
and dark matter halo models. We also detect the Monoceros stream as an
overdensity of ${\rm[Fe/H]}>-1.5$ dex stars in the $l=232\arcdeg$ and
$b=26\arcdeg$ direction at $R_{\rm gal}\lesssim25$ kpc. In the two sightlines
where we do not detect significant substructure, the median metallicity is found
to be independent of distance within systematic uncertainties
(${\rm[Fe/H]}\sim-1.5\pm0.1$ dex).
\end{abstract}


\section{Introduction}

Studies of the Galactic stellar halo set constraints on the formation history of
the Milky Way and galaxy formation processes in general. For example,
contemporary simulations of galaxy formation predict that stellar halos of
Milky-Way-type galaxies are assembled from inside out, with the majority of mass
($50\%-80\%$) coming from several massive ($10^8$$-$$10^{10}$ $M_\sun$)
satellites that have merged more than 9 Gyr ago, while the remaining mass comes
from lower mass satellites accreted in the past 5$-$9 Gyr \citep{bj05,dh08}.
The actual fraction of massive versus less massive mergers will depend on the
formation history of the galaxy in question.

A further prediction is that the stellar halos should be more centrally
concentrated and should have steeper density profiles at moderate radii
($>20-30$ kpc), than host dark matter halos (e.g., Figure 9 in \citealt{bj05} or
Figure 12 in \citealt{dh08}). According to \citet{bj05}, ``the difference in
profile shapes---and the steep rollover in the light matter at moderate to large
radii---is a natural consequence of embedding the light matter deep within the
dark matter satellites: the satellites' orbits can decay significantly before
any of the more tightly bound material is lost.'' Hence, they anticipate a
correlation between the extent of the stellar halo (steepness of the density
profile) and the extent (or mass) of satellites that built the halo: less
extended (more massive) satellites will build more concentrated stellar halos.
Therefore, the contribution of massive mergers to the formation of the Milky Way
halo can be constrained by characterizing the stellar halo number density
profile over large distances and over a wide sky area.

The large area covered by the Sloan Digital Sky Survey (SDSS; \citealt{york00}),
with accurate photometric measurements ($\sim0.02$ mag, \citealt{ivezic04}) and
faint flux limits ($r<22$), allowed for a novel approach to studies of the
stellar distribution in the Galaxy. Using a photometric parallax relation
appropriate for main-sequence stars, \citet[hereafter J08]{juric08} estimated
distances for a large number of stars and directly mapped the Galactic stellar
number density to heliocentric distances of 20 kpc. They found that the halo
stellar number density distribution within 20 kpc of the Sun can be fit with a
two parameter, single power-law ellipsoid model
\begin{equation}
\rho(R,Z) \propto [R^2 + (Z/q)^2]^{n/2}\label{oblate_halo},
\end{equation}
where $R$ and $Z$ are the cylindrical galactocentric radius and height above the
Galactic plane, respectively, $n=-2.77\pm0.2$ is the power law index, and
$q\equiv c/a = 0.64\pm0.1$ is the ratio of major axes in the $Z$ and $R$
direction, indicating that the halo is oblate (flattened in the $Z$ direction).
However, additional data suggest that the J08 single power law halo cannot be
extrapolated beyond 20 kpc. A kinematic analysis by Carollo et al.~(2007, 2010)
suggests that the halo consists of two components with different spatial density
profiles and median metallicities, with the ``inner'' to ``outer'' halo
transition happening at 15-20 kpc, and the density profile becoming
{\em shallower} beyond that point. On the other hand, the distribution of RR
Lyrae stars from the SEKBO survey \citep{keller08}, and RR Lyrae and
main-sequence stars from SDSS stripe 82 data seem to show a {\em steeper}
density profile beyond 30 kpc (main sequence stars can be detected up to 40 kpc
in coadded SDSS stripe 82; \citealt{sesar10}).

While these studies indicate a change in the halo density profile, each has its
shortcomings: kinematic studies do not give a direct measurement of the density
profile but rather model it under an assumed dark matter halo potential, RR
Lyrae stars are relatively sparse tracers of stellar number density
($\sim5$ kpc$^{-3}$ in the solar neighborhood; \citealt{sesar10} and references
therein), and the SDSS stripe 82 region covers only about $1\%$ of the sky.
Ideally, the halo stellar number density distribution should be mapped using
main-sequence stars over a large fraction of the sky and to a distance of at
least 50 kpc. Such studies will be enabled by next generation wide-area surveys
such as the Dark Energy Survey \citep{huan09}, Pan-STARRS \citep{kaiser02} and
Large Synoptic Survey Telescope (LSST; \citealt{ivezic08b, LSST}).

Meanwhile, the Canada-France-Hawaii Telescope Legacy Survey (CFHTLS) has
observed 170 deg$^2$ of sky in four fields as part of CFHTLS ``wide'' survey\footnote{\url{http://terapix.iap.fr/cplt/oldSite/Descart/summarycfhtlswide.html}}.
The names, positions, and sky coverage of CFHTLS ``Wide'' fields are listed in
Table~\ref{wide_fields}. Due to their relatively small sky coverage, these
fields are effectively ``pencil-beam'' surveys for the purposes of this paper.
Despite the small sky coverage (comparable to SDSS stripe 82), these data are
very useful because of their depth (95\% completeness at $i'=23.5$ for point
sources; \citealt{goranova09}), corresponding to a distance limit of $\sim35$
kpc for main-sequence stars, and because they observe lines of sight unexplored
by other surveys. These properties allow one to study how the halo density
profile changes as a function of distance and line of sight, both in the inner
and outer halo.

This paper is organized as follows. In Section~\ref{data}, we give an overview
of CFHTLS data and describe the synthesis of SDSS $ugri$ magnitudes from
recalibrated CFHTLS $u^* g^\prime r^\prime i^\prime i^\prime_2$ observations.
The recalibration and transformation of CFHTLS data into the SDSS photometric
system \citep{fukugita96} is done to allow the usage of CFHTLS data with
relations already defined on the SDSS system, such as the color-luminosity and
photometric metallicity relations \citep{ivezic08a,bond10}. The CFHTLS
$z^\prime$-band observations are not recalibrated as they were not publicly
available for all CFHTLS fields at the time of writing. In
Section~\ref{J08comparison}, we analyze the distribution of stellar counts as a
function of position and compare it to the J08 halo model. The best-fit broken
power-law model is derived in Section~\ref{sec.break}. In
Section~\ref{metallicity}, we study the metallicity distribution in the halo and
analyze the dependence of best-fit model parameters on the adopted metallicity
distribution. We finish by discussing our conclusions in
Section~\ref{discussion}.

\section{The Data\label{data}}

We perform several data quality tests and post-processing steps before using the
CFHTLS data in scientific analysis. First, we photometrically recalibrate the
data by utilizing repeated CFHTLS observations and then transform them to the
SDSS photometric system. We also investigate the performance of different types
of CFHTLS magnitudes.

\subsection{Overview of CFHTLS Data\label{CFHTLS_overview}}

We use the CFHTLS data processed by the {\em MegaPipe} image processing pipeline
\citep{gwyn08}. The pipeline takes as input MegaCam \citep{boulade03} images
detrended by the Elixir pipeline \citep{mc04}, and performs an astrometric and
photometric calibration on them. The calibrated images are resampled and
combined into image stacks. The catalogs of sources are derived by running
SExtractor \citep{ba96} on each image stack. The resulting catalogs only pertain
to a single band; no multi-band catalogs are generated by the MegaPipe pipeline.
The astrometry in these catalogs is accurate to within $0\arcsec.2$ relative to 
external reference frames \citep{gwyn08}. The catalogs are available for each
pointing (a $\sim1$ deg$^2$ part of a CFHTLS field) through the Canadian
Astronomy Data Centre (CADC) Web site\footnote{\url{http://www1.cadc-ccda.hia-iha.nrc-cnrc.gc.ca/community/CFHTLS-SG/docs/cfhtlswide.html}}.

Since multi-band catalogs are not available, we generate them for each pointing
by positionally matching sources detected in $u^* g^\prime r^\prime$ bands to
sources detected in the $i^\prime$ or $i^\prime_2$ band using a $1\arcsec.5$
matching radius. The $i^\prime$ and $i^\prime_2$ bands are the deepest of the
CFHT bands, so most objects detected in other bands will also be detected in
these two. The $i^\prime_2$ filter is the new CFHT filter that was installed
after the $i^\prime$ filter broke in 2007 October\footnote{See \url{http://www1.cadc-ccda.hia-iha.nrc-cnrc.gc.ca/megapipe/docs/filters.html}}.
We further match $\sim1.7$ million CFHTLS sources overlapping the SDSS footprint
to SDSS DR7 data \citep{abazajian09} using the $1\arcsec.5$ matching radius.

\subsection{Star-Galaxy Separation\label{SG}}

We separate point-like and extended sources using the half-light-radius (HLR)
measured (in pixel units) for each detected source by the SExtractor. For
point-like sources, the HLR is independent of magnitude and depends only on
image seeing \citep{schultheis06}. For each CFHTLS band and pointing, we remove
the dependence of HLR on image seeing by subtracting the median HLR value of
bright (17.5$-$18.5 mag) sources. Bright sources are used in this procedure
because they are dominated by point-like sources (see Figure 9 in
\citealt{gwyn08}). We find that after subtraction, the distribution of HLR
values of bright sources in the $r^\prime$ and $i^\prime$ (or $i^\prime_2$)
bands can be modeled as a $\sim0.1$ pixel wide Gaussian centered at zero.

We classify a source as a star if its HLR in the $r^\prime$ and $i^\prime$
(or $i^\prime_2$) bands is less than 0.2 pixels (after removing the dependence
of HLR on image seeing). Sources that do not satisfy this condition are
classified as galaxies. To estimate the quality of this classification, we
compare it to the SDSS star-galaxy classification \citep{lupton02} based on
deep, co-added SDSS stripe 82 data. The star-galaxy separation in co-added SDSS
stripe 82 data is reliable to at least $r\sim23$ (Annis, J. et al.(2011), in
preparation), and for purposes of this comparison we consider it to be the
ground truth.

Figure~\ref{classification} shows the fraction of SDSS stars identified as stars
in CFHTLS data (completeness) and the fraction of CFHTLS stars identified as
galaxies by the SDSS (contamination) as a function of $r^\prime$ magnitude.
Using this plot, we estimate that the observed number counts will be
underestimated by about $5\%$ for $r^\prime<21$ and  overestimated by about
15\%$-$20\% at the faint end ($r^\prime\sim22.5$).

\subsection{Recalibration and Transformation of CFHTLS Data to the SDSS $ugriz$
Photometric System\label{recalibration}}

Following \citet{pad08}, the photometric recalibration of CFHTLS data is
separated into ``relative'' and ``absolute'' calibration. This process is
schematically illustrated in Figure~\ref{recal_schema}. For relative
calibration, we use overlaps between pointings to recalibrate all pointings
within a CFHTLS field onto a single photometric system (specific to that field).
This calibration is then tied to the SDSS photometric system (absolute
calibration). The final step in the process is the synthesis of SDSS $ugri$
photometry from recalibrated CFHTLS $u^*g^\prime r^\prime i^\prime i^\prime_2$
observations.

\subsubsection{Flux Extraction\label{flux_extraction}}

The CFHTLS catalogs provide adaptive-aperture ($m_{auto}$) and fixed-aperture
magnitudes ($m_{aper}$) for sources detected by SExtractor. To compare fluxes
extracted by these two methods, we select stars from the W3 field and calculate
$m_{auto}-m_{aper}$ residuals, where $m$ stands for $u^*$, $g^\prime$,
$r^\prime$, and $i^\prime$ bands. The residuals are binned in $m_{auto}$ bins
and median values are calculated for each bin. The dependence of median values
on $m_{auto}$ for $u^*g^\prime r^\prime i^\prime$ bands is shown in
Figure~\ref{fig1}. In general, we find that the median difference between
adaptive- and fixed-aperture magnitudes increases linearly towards fainter
magnitudes, reaching $\sim0.04$ mag in the $i^\prime$ band.

This indicates that there is a problem with one or both extraction methods.
As we will demonstrate in Section~\ref{photometry_check}, the adaptive-aperture
magnitudes ($m_{auto}$) are responsible for the dependence seen in
Figure~\ref{fig1}, based on the following reasons. First, measuring
adaptive-aperture magnitudes is a more complex process than measuring at the
fixed aperture (for details see SExtractor v2.5 manual\footnote{\url{https://www.astromatic.net/pubsvn/software/sextractor/trunk/doc/sextractor.pdf}}), making
adaptive-aperture magnitudes less robust. Second, a caveat in the SExtractor
v2.5 manual warns of potential problems with adaptive-aperture magnitudes when
the SExtractor $R_{min}$ parameter is set too low. This caveat states that,
``when signal to noise is low, it may appear that an erroneously small aperture
is taken by the algorithm. That is why we have to bound the smallest accessible
aperture to $R_{min}$''. Therefore, if $R_{min}$ was set too low during flux
extraction, inadequate (too small) apertures may have been used, making
$m_{auto}$ fainter and causing $m_{auto}-m_{aper}$ residuals to be biased
towards positive values. Even though the caveat states that this may happen at
fainter magnitudes (low signal-to-noise), Figure~\ref{fig1} seems to indicate
that brighter magnitudes may be affected as well. We therefore use
fixed-aperture magnitudes in the rest of this work.

\subsubsection{Relative Calibration\label{rel_cal}}

Relative calibration places all pointings within a CFHTLS field onto a single
photometric system (specific to that field) using repeatedly observed stars that
are located in regions where pointings overlap (CFHTLS pointings overlap about
$3\arcsec$ in right ascension and $4\arcsec$ in declination directions).
Mathematically, the problem of relative calibration can be described as a
$\chi^2$ minimization problem \citep{pad08} with
\begin{equation}
\chi^2 = \sum^{n_{star}}_i \chi^2_i = \sum^{n_{star}}_i \sum_{j\in \mathcal{O}(i)} \left [ \frac{m_i - m_{j,orig} - \Delta^{rel}_{j,f}}{\sigma_j} \right ]^2 \label{chi},
\end{equation}
where $n_{star}$ is the number of unique, repeatedly observed stars in the
$f=$ W1, W2, W3, W4 field, $j$ runs over multiple observations (pointings),
$\mathcal{O}(i)$, of the $i$th star with unknown true magnitude $m_i$,
$m_{orig}$ and $\sigma$ are the magnitude and its error supplied by MegaPipe,
respectively, and $\Delta^{rel}_{j,f}$ is the correction needed to place the
$j$th pointing on the $f$ field's relative photometric system. The system
defined with Equation~\ref{chi} is overdetermined, since the number of unknowns
($n_{star}+n$[parameters]) is smaller than the number of observations
($n_{obs}$ is at least $2n_{star}$). Equation~\ref{chi} can be expressed in
matrix form and can be solved using sparse matrix techniques (see Section 3.2 in
\citealt{pad08}).

To solve Equation~\ref{chi} for each field, we use repeatedly observed stars
with $17 < m_{orig} < 23$, where
$m_{orig}=u^*,g^\prime,r^\prime,i^\prime,i^\prime_2$. The observations in this
magnitude range are not saturated and their reported photometric errors are
smaller than 0.2 mag. The observations are supplied to a sparse matrix
inversion code written by \citet{pad08}, and the best-fit $\Delta^{rel}_{j,f}$
values returned by this code are renormalized so that the median value of
$\Delta^{rel}_{j,f}$ is equal to zero. On average, we find the rms scatter of
$\Delta^{rel}_{j,f}$ values to be between 0.02 and 0.03 mag, reflecting the
systematic uncertainty in CFHTLS magnitudes.

\subsubsection{Absolute Calibration\label{abs_cal}}

Following relative calibration, we tie relative photometric systems to the SDSS
photometric system using CFHTLS stars matched to SDSS stars. The goal is to find
the best-fit $\Delta^{abs}_f$ value for each field, such that
\begin{equation}
\chi^2 = \sum_{f=W1}^{W4} \sum_{i} \left[ \frac{m_{i,rel}-m_{i,sdss}-C\, color - \Delta^{abs}_f}{\sqrt{\sigma_i^2+\sigma_{i, sdss}^2}} \right]^2
\end{equation}
is minimized, where $i$ runs over CFHTLS stars in the $f$ field matched to SDSS
stars, $m_{rel}$ is the CFHTLS magnitude after relative calibration
($m_{rel} = m_{orig} + \Delta^{rel}_{j,f}$), and $m_{sdss}$ and
$\sigma_{sdss}$ are SDSS point-spread function (PSF) magnitude and its error,
respectively. For $m_{rel}=u^*,g^\prime,r^\prime,i^\prime,i^\prime_2$, the SDSS
color and band are color$=u-g,g-r,g-r,r-i,r-i$ and $m_{sdss}=u,g,r,i,i$,
respectively (colors and magnitudes are {\em not} corrected for interstellar
medium (ISM) extinction).

The $C$ color term corrects the linear dependence of $m_{i,rel}-m_{i,sdss}$
residuals on color caused by differences between CFHTLS and SDSS spectral
response curves (see Figure 1 in \citealt{gwyn08} for a comparison of spectral
response curves). This term reduces the scatter in $m_{i,rel}-m_{i,sdss}$
residuals and improves our estimate of $\Delta^{abs}_f$. Since CFHTLS and SDSS
spectral response curves should not change significantly with time, the
$C$ color term is simultaneously fit for all four CFHTLS fields. Note that
this term is only used when estimating $\Delta^{abs}_f$; linear and other
higher-order color terms that model the transformation of CFHTLS bands into
SDSS bands are derived in Section~\ref{transformation}.

To determine $\Delta^{abs}_f$ values, we use CFHTLS stars with $m_{rel}>17$
matched to SDSS stars with $m_{sdss}<21$ and with the SDSS color as specified in
Table~\ref{abs_coeff}. The best-fit $C$ and $\Delta^{abs}_f$ are also listed in
Table~\ref{abs_coeff}. Finally, we define recalibrated CFHTLS magnitudes as
\begin{equation}
m_{cfht} = m_{j,f,orig} + \Delta^{rel}_{j,f} + \Delta^{abs}_f = m_{j,f,orig} + \Delta^{tot}_{j,f},\label{m_cfht}
\end{equation}
where $\Delta^{tot}_{j,f}$ is the total correction for the $j$th pointing in the
$f$ field. The $\Delta^{tot}_{j,f}$ values for CFHTLS fields are listed in
Tables~\ref{W1_total_coeff}.

\subsubsection{Transformation to SDSS Bandpasses\label{transformation}}

With CFHTLS $m_{cfht}=u^*,g^\prime,r^\prime,i^\prime,i^\prime_2$ observations
calibrated onto the SDSS system, we now derive the equations that transform
CFHTLS observations into SDSS $m_{sdss}=u,g,r,i,i$ magnitudes. In general, the
transformation from CFHTLS to SDSS bandpasses can be defined as
\begin{equation}
m = m_{cfht} + f(color) + Z_0\label{cfht2sdss},
\end{equation}
where $Z_0$ is the constant term, and $f(color)$ is some function of CFHT colors
(colors and magnitudes are {\em not} corrected for ISM extinction).

To find $f(color)$ and $Z_0$ for each SDSS $ugri$ bandpass, we bin
$m_{sdss}-m_{cfht}$ residuals as a function of CFHT color, and fit polynomials
to $m_{sdss}-m_{cfht}$ medians. Here we only use CFHTLS stars with $m_{cfht}>17$
matched to SDSS stars with $m_{sdss}<20$. The dependence of $m_{sdss}-m_{cfht}$
residuals on CFHT color is shown in Figure~\ref{cfht2sdss_plot}.

The best-fit polynomials shown in Figure~\ref{cfht2sdss_plot} define the
transformation of recalibrated CFHTLS magnitudes into SDSS magnitudes:
\begin{eqnarray}
u &=& u^* + 1.07 - 2.375(u^*-g^\prime) + 1.954(u^*-g^\prime)^2 - 0.483(u^*-g^\prime)^3\label{synth_u} \\
g &=& g^\prime + 0.05 - 0.062(g^\prime-r^\prime) + 0.365(g^\prime-r^\prime)^2 \\
r &=& r^\prime - 0.05 + 0.275(g^\prime-r^\prime) - 0.380(g^\prime-r^\prime)^2 \\
i &=& i^\prime - 0.002 + 0.092(r^\prime-i^\prime) - 0.015(r^\prime-i^\prime)^2 \\
i &=& i^\prime_2 - 0.005 + 0.145(r^\prime-i^\prime_2) - 0.280(r^\prime-i^\prime_2)^2 + 0.140(r^\prime-i^\prime_2)^3\label{synth_i}.
\end{eqnarray}
Using Equations~\ref{synth_u} to~\ref{synth_i}, we synthesize SDSS photometry
($m=u,g,r,i$) for all CFHTLS observations. Note that these transformations can
be used even if the relative recalibration step from Section~\ref{rel_cal} is
skipped because the best-fit $\Delta^{rel}_{j,f}$ values are renormalized so
that their median is zero.

\subsection{Quality of Photometric Calibration\label{photometry_check}}

To estimate the quality of synthesized photometry, we use CFHT sources brighter
than $m=17$ that are matched to SDSS stars with $m_{sdss}<20$, where $m$ stands
for $u$, $g$, $r$, and $i$. For each band $m$ we calculate $m-m_{sdss}$
residuals, and bin them as a function of color (not corrected for ISM
extinction, color$=u-g,g-r,g-r,r-i$ for $m=u,g,r,i$, respectively). The absolute
value of $m-m_{sdss}$ medians for $m=g,r,i$ bands is smaller than 0.01 mag and
shows no dependence on color, for all CFHTLS fields. For W1, W3, and W4 fields,
the absolute value of $u-u_{sdss}$ medians is smaller than 0.02 mag, and shows
no dependence on $u-g$ color, while for the W2 field the $u-u_{sdss}$ medians
seem to show linear dependence on $u-g$ color for $u-g<1.3$, as illustrated in
Figure~\ref{color_check} (bottom panel). Since all four fields were calibrated
using the same procedure, and the transformations from recalibrated CFHTLS to
SDSS magnitudes were derived using the data from all fields, the $u-u_{sdss}$
dependence on $u-g$ color points to a problem with the original CFHTLS
$u^*$-band observations in the W2 field. We hypothesize that the $u-u_{sdss}$
dependence on color may be due to incorrectly determined color-dependent
air-mass term by Elixir or MegaPipe pipelines.

The systematic uncertainty of synthesized magnitudes can be determined by
comparing synthesized photometry of repeatedly observed CFHT sources; such
sources can be found in regions where pointings overlap. We find that the
systematic uncertainty is $\sim0.03$ mag (see Figure~\ref{error_vs_mag}),
which is consistent with the value cited by \citet{gwyn08}.

To quantify the non-linearity of synthesized photometry, we bin $m-m_{sdss}$
residuals in $m=u,g,r,i$ magnitude bins. As shown in
Figure~\ref{auto_aper_sdss_comp}, the medians of $g-g_{sdss}$ residuals in $g$
magnitude bins show no dependence on magnitude, indicating linear behavior of
synthesized photometry. We have repeated this last test using adaptive-aperture
magnitudes and have found a magnitude dependence in $g-g_{sdss}$ residuals
similar to the one shown in Figure~\ref{fig1}. Similar results are also obtained
for $uri$ bands. These results point to the dependence shown in
Figure~\ref{fig1} as being due to the problematic adaptive-aperture flux
extraction, and justify our choice of fixed-aperture magnitudes.

\section{Analysis of the Number Density Distribution Profiles\label{J08comparison}}

The data presented in previous sections allow us to measure the stellar number
density of near main sequence turnoff (MSTO) halo stars, and examine it as a
function of position in the Galaxy. The sample we obtained from the CFHTLS
extends to distances and Galactocentric radii ($R_{gal}$) nearly a factor of two
greater than previous wide-area studies that used main-sequence stars (e.g.,
$5 \lesssim R_{gal} \lesssim 15$ kpc; J08). The sample also overlaps in range
with studies based on RR Lyrae stars ($5 \lesssim R_{gal} \lesssim 110$ kpc;
\citealt{sesar10}), and it therefore presents an opportunity to examine the
behavior of the halo density profile in the intermediate range
($5 \lesssim R_{gal} \lesssim 30$ kpc).

We begin by selecting a sample of near-MSTO stars with the following criteria:
\begin{eqnarray}
0.2 < g-r < 0.3 \\
g > 17\, \, \& \, \, 17 < r < 22.5\, \, \& \, \, i > 17 \\
5 < D/{\rm kpc} < 35
\end{eqnarray}
where magnitudes and colors are corrected for ISM extinction using maps from
\citet{SFD98}. The $g-r$ color cut serves to select near-MSTO stars, and $D$ is
the heliocentric distance to a star computed using the photometric-parallax
relation from \citet[see Equations A2 and A7 in their Appendix A]{ivezic08a}:
\begin{equation}
M_r = -0.56 + 14.32x - 12.97x^2 + 6.127x^3 - 1.267x^4 + 0.0967x^5 - 1.11[Fe/H] - 0.18[Fe/H]^2. \label{M_r}
\end{equation}
The $x=g-i$ color in Equation~\ref{M_r} is computed from the more accurately
measured $g-r$ color using the stellar locus fit from J08 (see their Figure 8).
The $[Fe/H]$ is estimated from the $u-g$ and $g-r$ color using the photometric
metallicity relation from \citet{bond10} (see their Equation A1)
\begin{equation}
[Fe/H] = -13.13 + 14.09x + 28.04y -5.51xy -5.90x^2 -58.68y^2 + 9.14x^2y -20.61xy^2 + 58.20y^3, \label{photo_FeH}
\end{equation}
where $x=u-g$ and $y=g-r$.
After these cuts, 13692, 7347, 6505 and 6676 stars are left in W1, W2, W3, and
W4 beams, respectively.

For each pencil beam, we bin the resulting subset in $\Delta DM = 0.2$~mag wide
bins in distance modulus\footnote{Binning in bins of equal size in distance
modulus (as opposed to distance) results in approximately equal number of stars
per bin, a consequence of the halo density profile being close to $R_{gal}^{-3}$
power law.}, $DM = 5\log(D)-5$. We thus obtain the distribution of number counts
$\Delta N(l, b, D)$, as a function of distance $D$ for each pencil beam.

We transform the observed counts to density:
\begin{equation}
\rho(l_i, b_i, D) = \frac{\Delta{}N(l_i, b_i, DM)}{0.2 \ln(10) D^3 \Delta\Omega \Delta{}DM}
\end{equation}
where $l_i, b_i$ and $\Delta\Omega_i$ are the field centers and area covered by
each beam as listed in Table~\ref{wide_fields}, and $\rho(l_i, b_i, D)$ is the
number density in stars~pc$^{-3}$. The sightlines sampled by CFHTLS data are
shown in Figure~\ref{xsections}.

In panels of Figure~\ref{profiles.r.J08}, we plot the dependence of measured
number density on the distance from the Galactic center, $R_{gal}$ for each of
the four CFHTLS beams. The measurements are marked by symbols with error bars
and connected with a solid line for clarity. Overplotted with a dotted line on
each panel is the prediction of the axisymmetric oblate halo model of J08.
Plotted as open circles, and connected by red line segments, are samples within
$|Z| \leq 5$ kpc of the Galactic plane. As these may be contaminated by disk
stars, we leave them out of all further analyses. The residuals of the data for
the J08 model are shown in Figure~\ref{profiles.r.diff.SJI10}.

The density profile observed in W1 beam clearly stands out. While it roughly
(within $\sim 20$\%) agrees with the predictions of J08 at $R_{gal}\sim 15$ kpc,
beyond that radius the observed density begins to exceed the J08 extrapolation,
peaking with a factor of $\sim2$ excess at $R_{gal} \sim 28$ kpc, and dropping
towards the end of the observed range ($R_{gal} \sim 35$ kpc). By comparing the
location of this overdensity with the best-fit model of the Sagittarius dwarf
spheroidal galaxy and its tidal tails \citep{lm10}, we conclude that the excess
is due to the leading and trailing arm of the Sagittarius stream
\citep{igi94, maj03} crossing the W1 beam. While the distribution and
metallicity of stars in this beam may provide useful new constraints for the
study of the stream (see Section~\ref{discussion}), this (un)fortunate fact
makes the majority of W1 data unusable for the study of the smooth halo profile.

Density profiles in W3 and W4 beams agree within $\sim 15$\% with the
predictions of the J08 halo model in $10 < R_{gal}/{\rm kpc} < 25$ and
$8 < R_{gal}/{\rm kpc} < 27$ ranges, respectively. The observed agreement is
nontrivial. First, the directions observed by these beams do not overlap with
the SDSS data used by J08, and therefore test their model in an entirely
different part of the halo (especially W4 beam). Second, the CFHTLS data cover
a significantly larger distance range than the data used by J08, thus validating
the extrapolation to $\sim10$ kpc greater distances (and an order of magnitude
change in stellar number density). And finally, while J08 did use a small
($\sim 300$ deg$^2$) area in the southern Galactic hemisphere to construct their
model, their best-fit parameters were largely determined by the $\gtrsim 6000$
deg$^2$ of data from the north. The fact that W4 beam ($b=-41\arcdeg.84$) is
very well matched by the model puts a constraint to any asymmetries between the
northern and southern hemispheres to $ \lesssim 15$\%, out to distances of at
least $R_{gal} \sim25$ kpc. This conclusion is also supported to distances of
$D \lesssim 18$ kpc by the analysis of north versus south SDSS III wide-area
imaging (Bonaca, A. et al.(2011), in preparation).

\begin{figure}
\epsscale{1.0}
\plotone{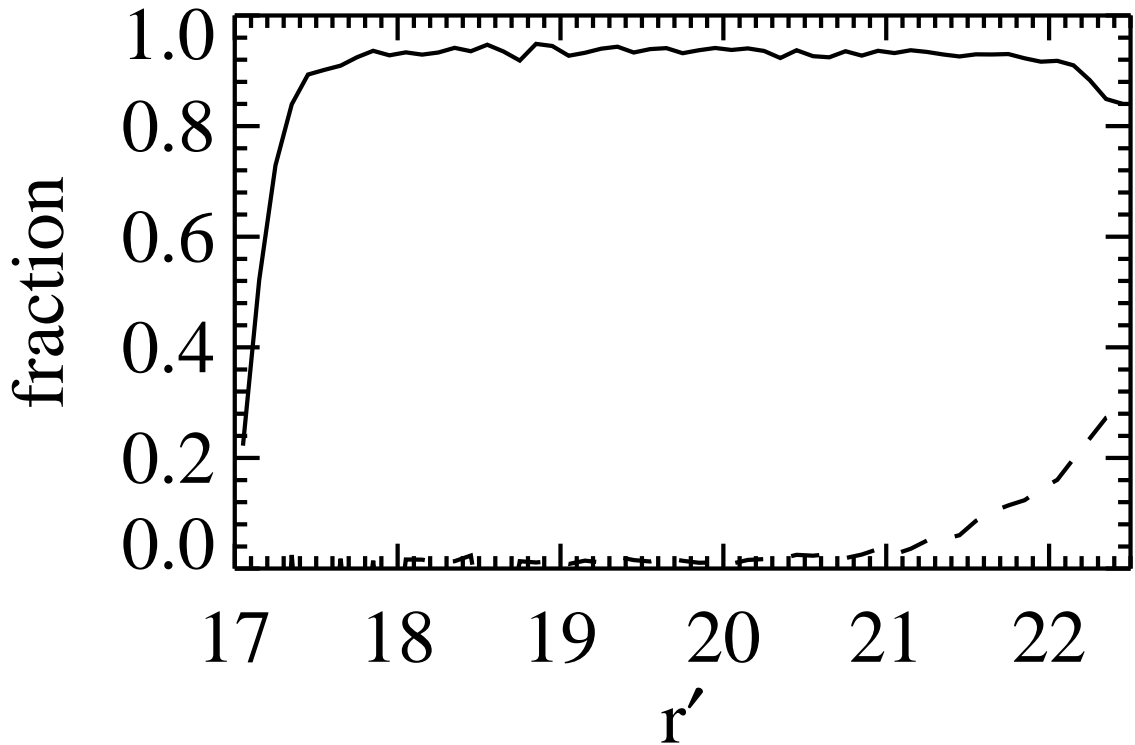}
\caption{
The fraction of SDSS stars identified as stars in CFHTLS data (completeness,
solid line) and the fraction of CFHTLS stars identified as galaxies by the SDSS
(contamination, dashed line) as a function of CFHTLS $r^\prime$ magnitude (not
corrected for ISM extinction). Using this plot, we estimate that the observed
number counts will be underestimated by about $5\%$ for $r^\prime<21$ and
overestimated by about 15\%$-$20\% at the faint end ($r^\prime\sim22.5$).
\label{classification}}
\end{figure}

\begin{figure}
\epsscale{1.0}
\plotone{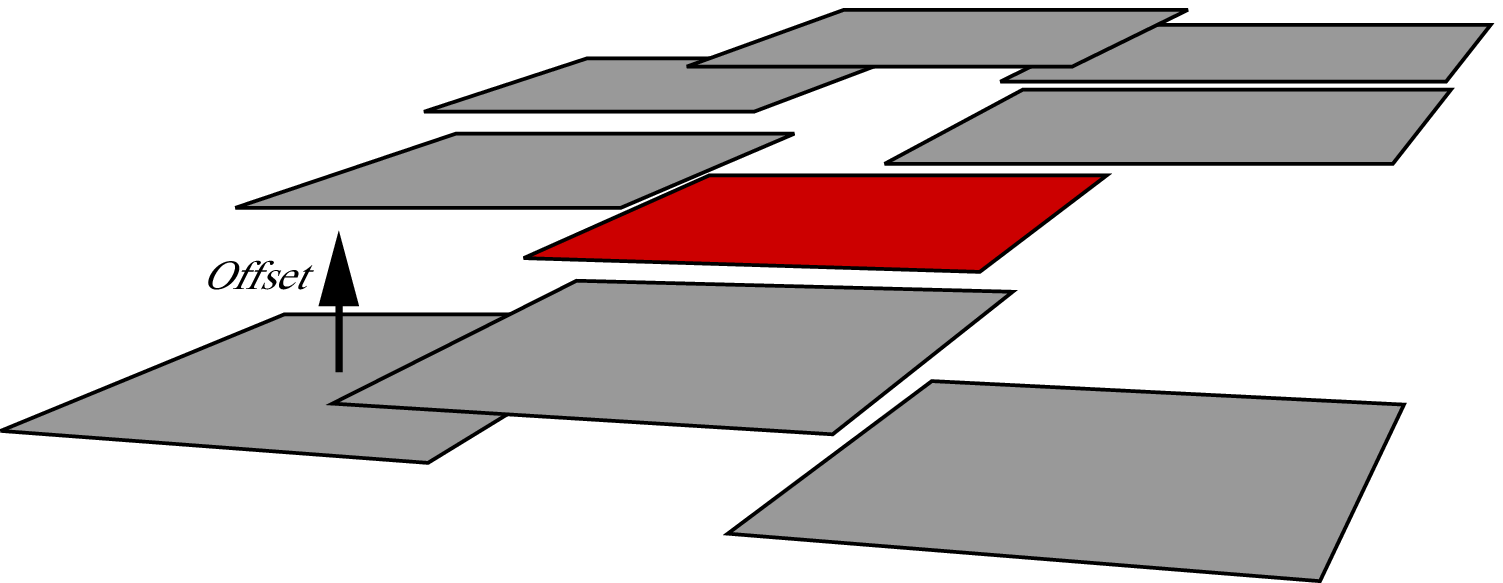}

\plotone{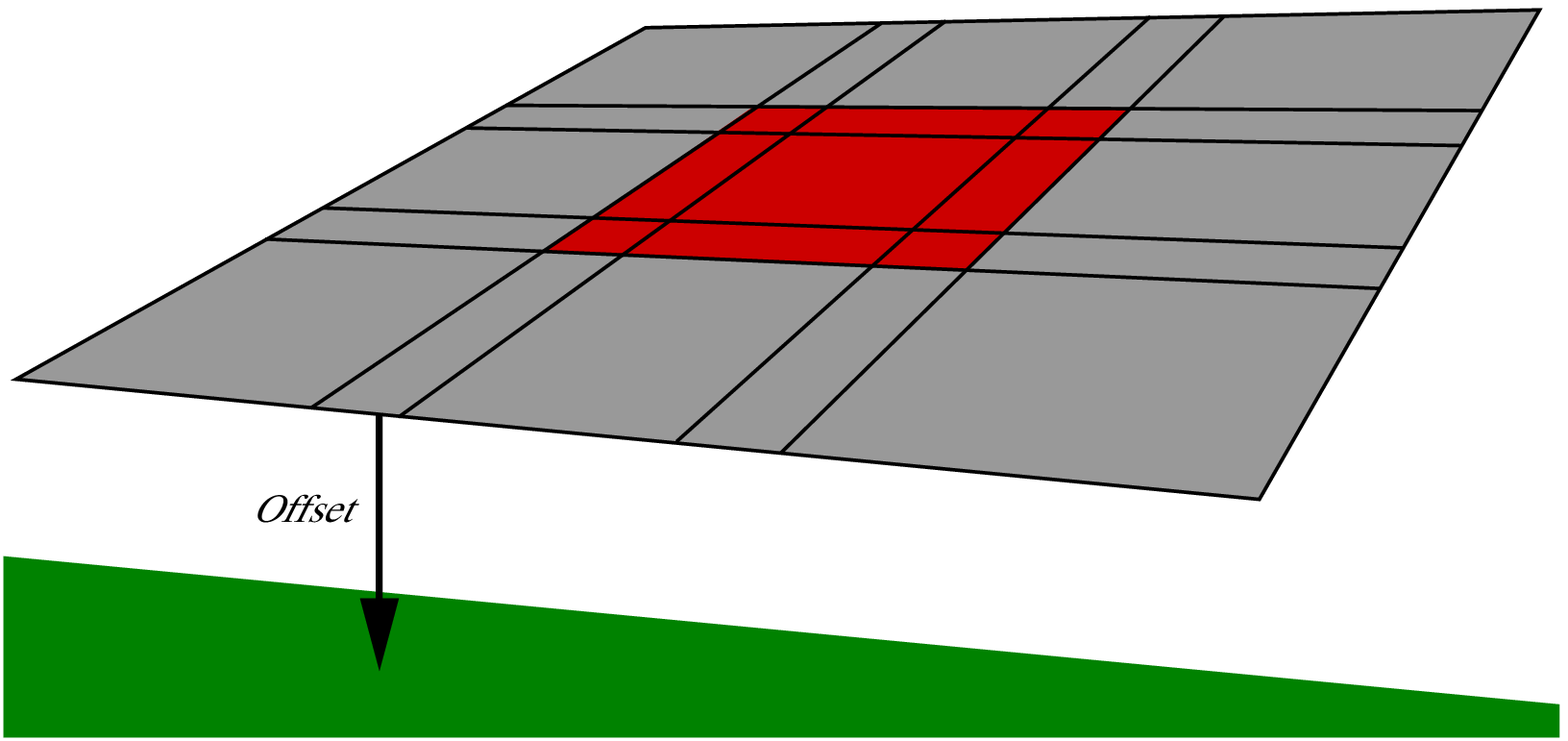}
\caption{
Schematic of the calibration process. For relative calibration ({\em top}),
the pointings within a CFHTLS field (9 {\em tiles}) are recalibrated to a
photometric system specific to that field using $\Delta_j^{rel}$ offsets (see
Section~\ref{rel_cal} for details). For absolute calibration ({\em bottom}), the
field is calibrated to the SDSS system ({\em light/green plane}) using a
$\Delta_f^{abs}$ offset (see Section~\ref{abs_cal} for details).
\label{recal_schema}}
\end{figure}

\begin{figure}
\epsscale{1.0}
\plotone{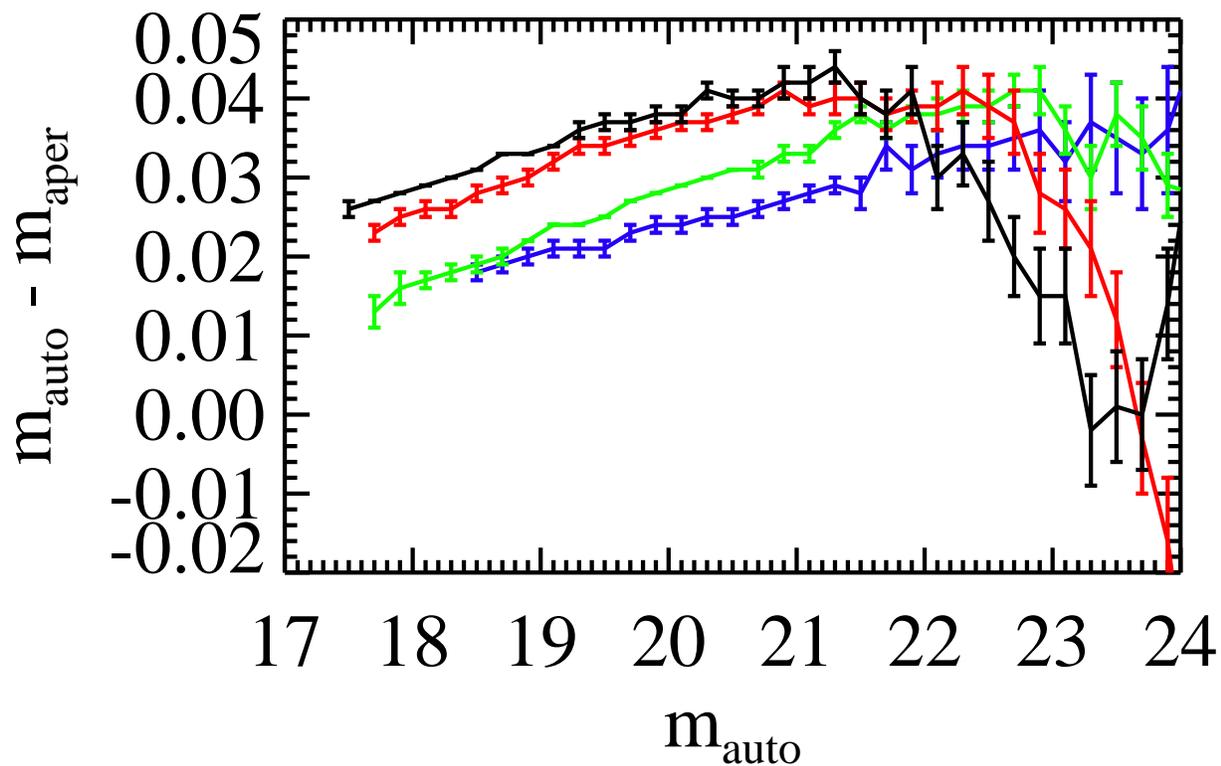}
\caption{
Dependence of $m_{auto}-m_{aper}$ residuals of stars on $m_{auto}$, where
$m_{auto}$ are adaptive-aperture magnitudes, $m_{aper}$ are fixed-aperture
magnitudes, and $m$ stands for $u^*$, $g^\prime$, $r^\prime$, and $i^\prime$
bands ({\em blue}, {\em green}, {\em red}, and {\em black} lines, respectively).
\label{fig1}}
\end{figure}

\begin{figure}
\epsscale{1.0}
\plotone{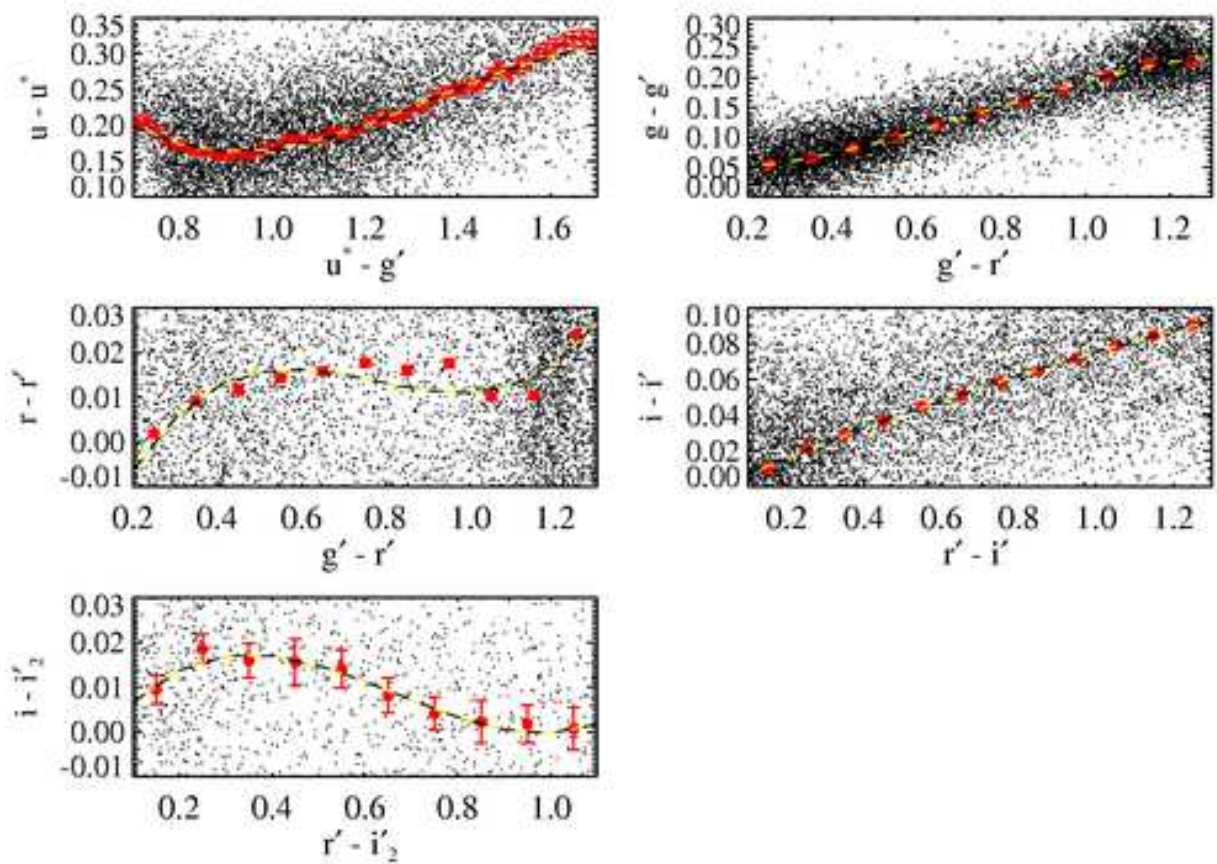}
\caption{
Dependence of $m_{sdss}-m_{cfht}$ residuals on CFHT color, where
$m_{sdss}=u,g,r,i,i$ are SDSS PSF magnitudes and
$m_{cfht}=u^*,g^\prime,r^\prime,i^\prime,i^\prime_2$ are recalibrated CFHTLS
magnitudes. The $m_{sdss}-m_{cfht}$ residuals are shown as dots (not all are
shown for clarity) and the symbols show their median values in color bins. The
error bars show errors in medians. The dashed lines were obtained by fitting
polynomials to $m_{sdss}-m_{cfht}$ medians.
\label{cfht2sdss_plot}}
\end{figure}

\begin{figure}
\epsscale{0.6}
\plotone{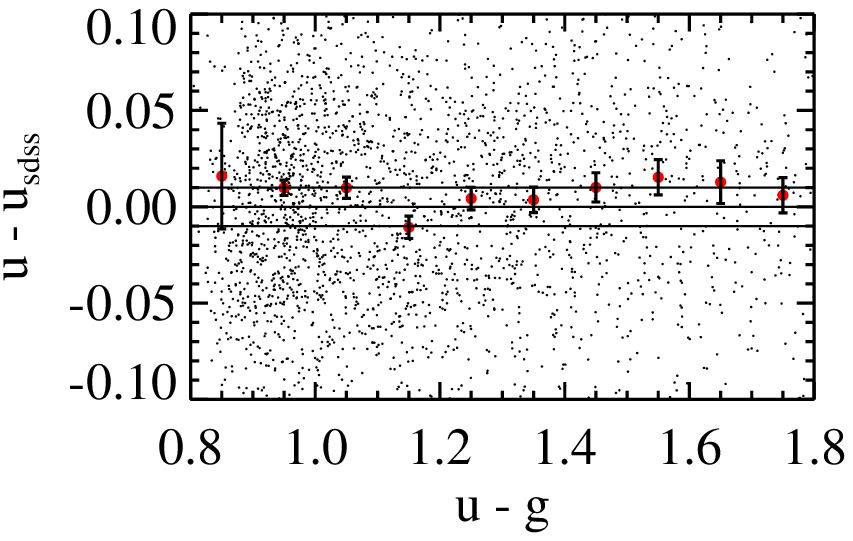}

\plotone{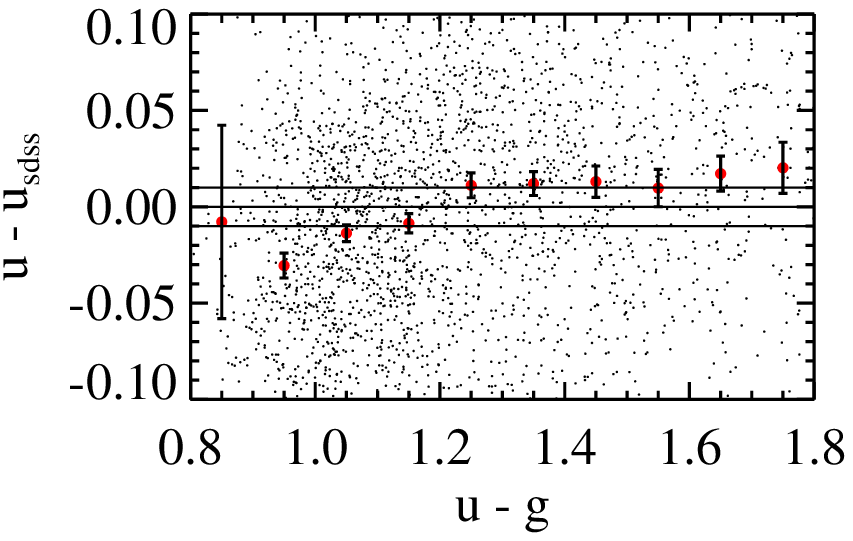}
\caption{
Dependence of $u-u_{sdss}$ residuals ({\em dots}) on synthetic $u-g$ color
(not corrected for ISM extinction) for the W1 ({\em top}) and W2 ({\em bottom})
fields, where $u$ and $g$ are synthetic observations derived from recalibrated
CFHTLS observations, and $u_{sdss}$ is the SDSS PSF $u$ band magnitude. The
symbols show $u-u_{sdss}$ medians in $u-g$ color bins, and error bars show the
error in median. To guide the eye, the solid lines show $u-u_{sdss} = \pm 0.01$
mag. For the W1 field, the $u-u_{sdss}$ medians are within 0.02 mag and do not
depend on $u-g$ color, while for the W2 field the $u-u_{sdss}$ medians seem to
show linear dependence on $u-g$ color for $u-g<1.3$, indicating a possible
problem with CFHTLS $u^*$ band observations in the W2 field.
\label{color_check}}
\end{figure}

\begin{figure}
\epsscale{1.0}
\plotone{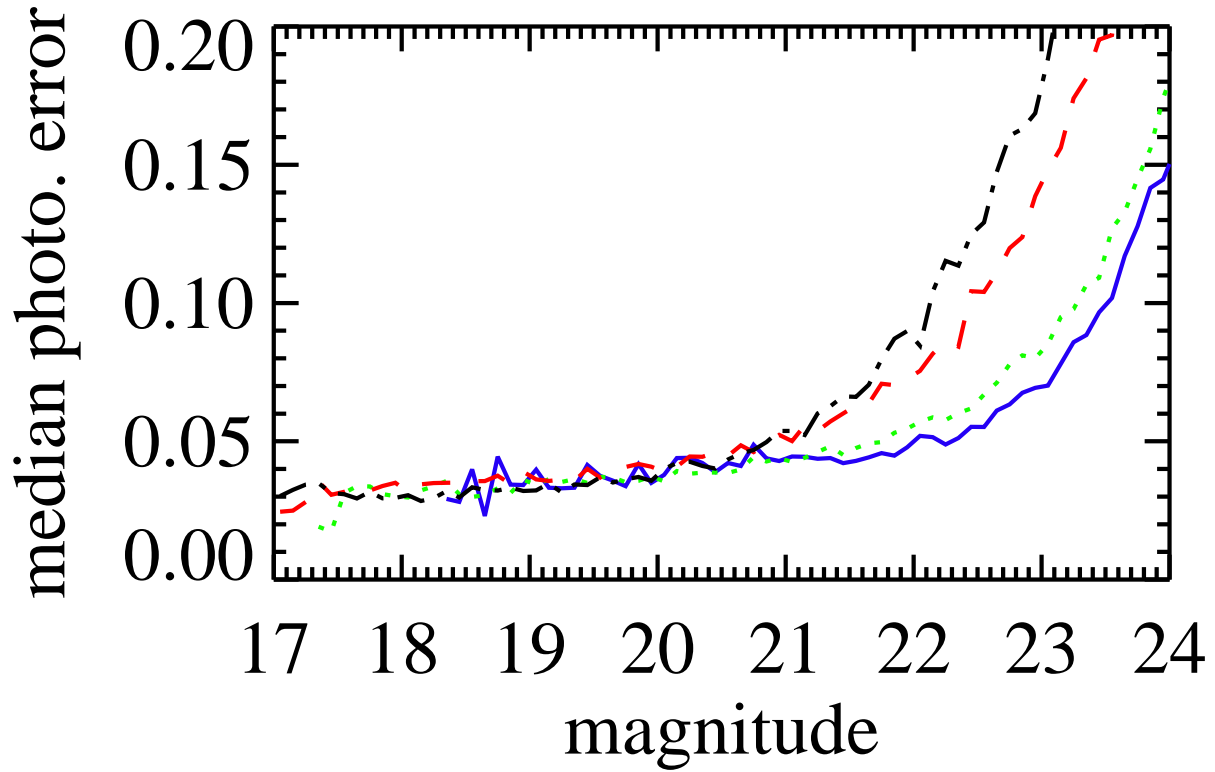}
\caption{
Median photometric error as a function of magnitude for synthetic $u$-band
({\em solid}), $g$-band ({\em dotted}), $r$-band ({\em dashed}), and $i$-band
observations ({\em dot-dashed}). The median photometric error was calculated as
the rms scatter of $m_2 - m_1$ residuals in magnitude bins, where $m_1$ and
$m_2$ are repeated observations of a star. The systematic uncertainty in
synthetic $ugri$ magnitudes is $\sim0.03$ mag, as indicated by the median
photometric error at the bright end (magnitudes brighter than $\sim20$ mag).
\label{error_vs_mag}}
\end{figure}

\begin{figure}
\epsscale{1.0}
\plotone{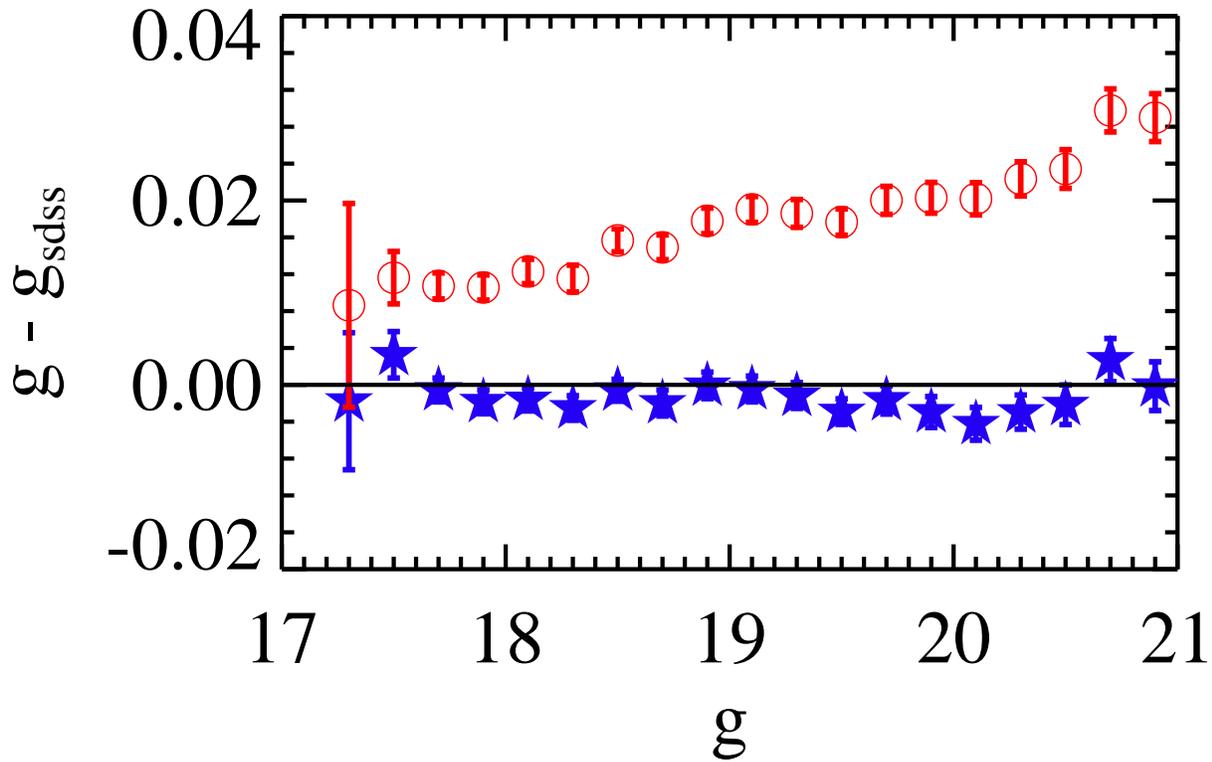}
\caption{
Dependence of median $g-g_{sdss}$ residuals on $g$ magnitude, where $g$ are
recalibrated fixed- ({\em stars}) and adaptive-aperture CFHTLS magnitudes
({\em open circles}), and $g_{sdss}$ is the PSF magnitude measured by SDSS. The
error bars indicate errors in medians. This comparison of CFHTLS and SDSS
magnitudes shows that the behavior seen in Figure~\ref{fig1} is due to
incorrectly measured adaptive-aperture magnitudes. Similar results are obtained
for $u$, $r$, and $i$ magnitudes.
\label{auto_aper_sdss_comp}}
\end{figure}

\begin{figure}
\epsscale{1.0}
\plotone{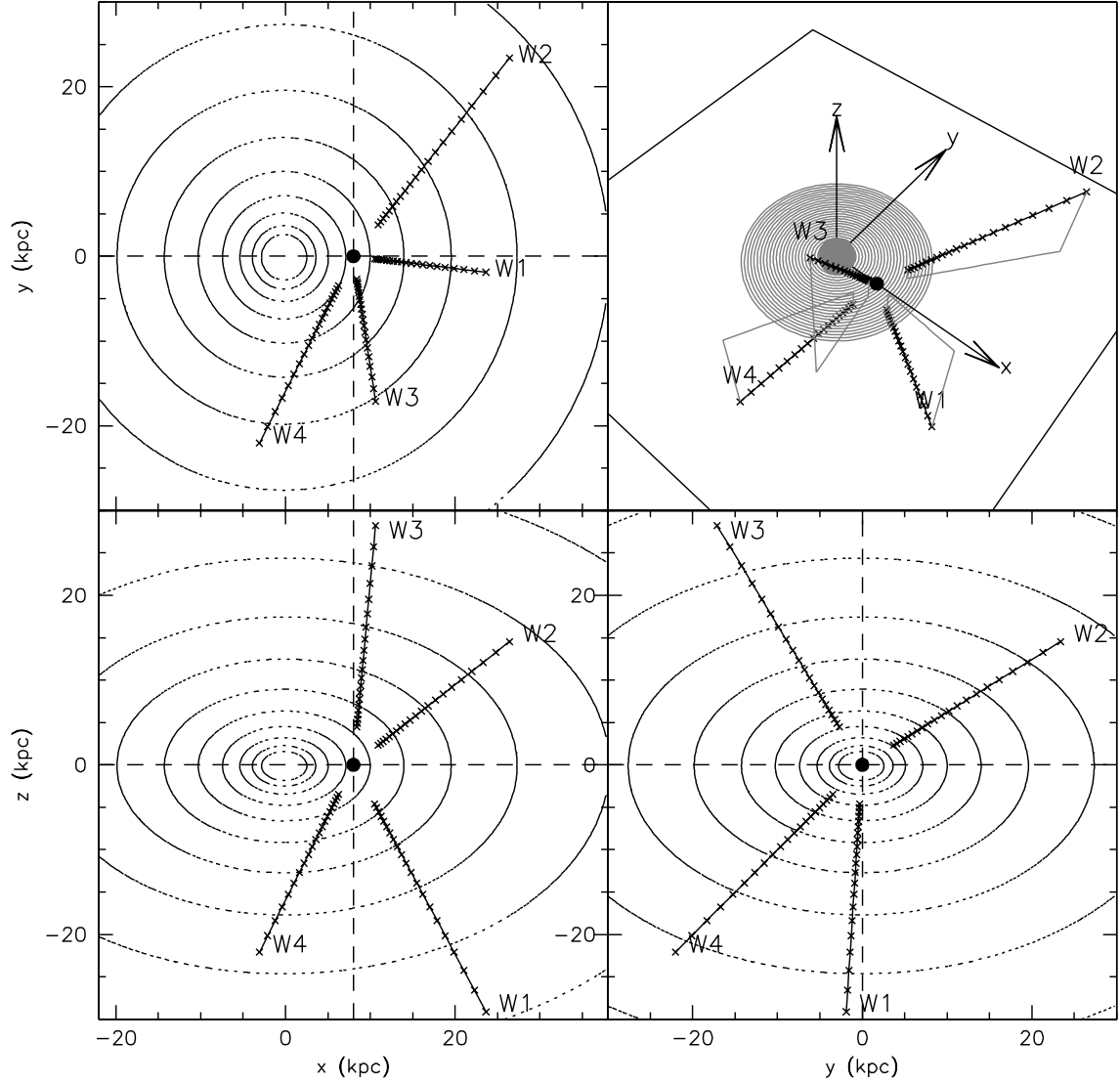}
\caption{
Visualization of the geometry of CFHTLS wide survey beams used in this paper,
overplotted on isodensity contours of the J08 halo.
\label{xsections}}
\end{figure}

\begin{figure}
\epsscale{0.8}
\plotone{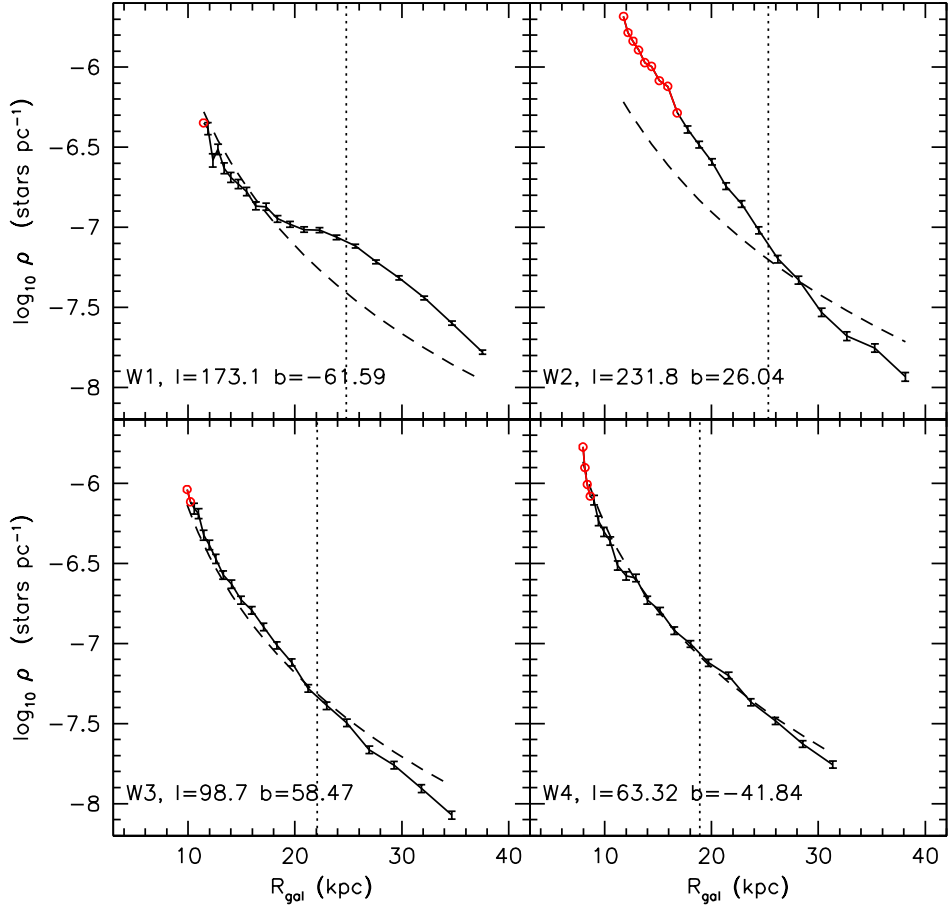}
\caption{
Stellar number density, measured in four CFHTLS wide-area survey beams as a
function of distance from the Galactic center, $R_{gal}$. Open circles denote
the measurements within 5 kpc of the Galactic plane, where the contamination by
disk stars may be greater than 10\%. For clarity, the symbols have been
connected by solid lines. Overplotted as a dashed line is the oblate power law
halo model from J08. Its overall normalization has been adjusted to fit the W3
and W4 data at $R_{gal} < 25$ kpc, as well as W1 data points satisfying
$R_{gal} < 15$ kpc (to avoid contamination by the Sagittarius stream). The
vertical line shows the J08 distance limit. The excess density at
$R_{\rm gal} \gtrsim 15$ kpc in the W1 field can be associated with the
Sagittarius stream, while the overdensity at $R_{\rm gal} \lesssim 25$ kpc in
the W2 beam is consistent with the location of the Monoceros stream.
\label{profiles.r.J08}}
\end{figure}

\begin{figure}
\epsscale{1.0}
\plotone{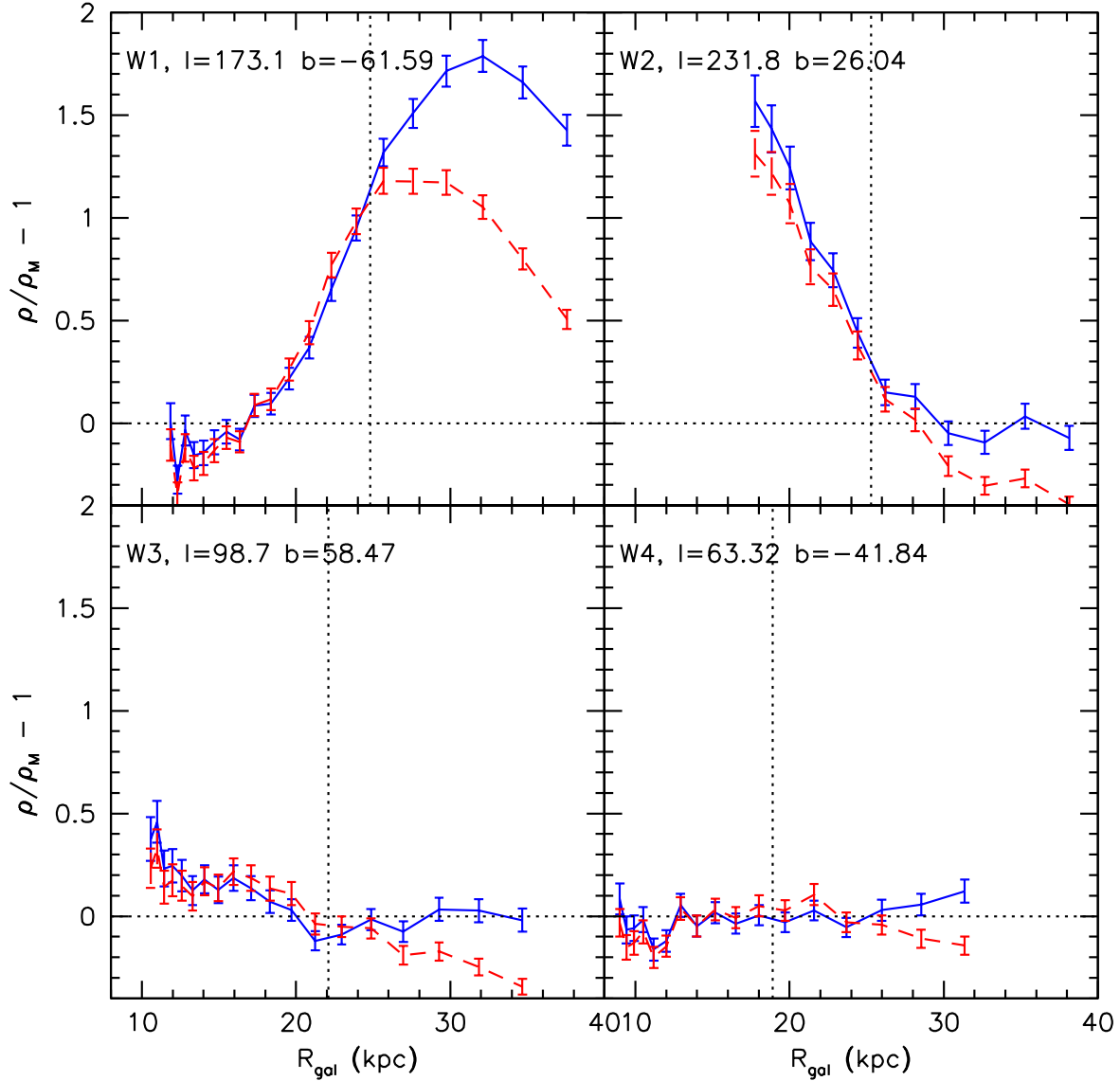}
\caption{
${\rm Data}/{\rm model} - 1$ residuals of the J08 model presented in
Figure~\ref{profiles.r.J08} ({\em dashed line}) and of the broken power law
model discussed in Section~\ref{sec.break} ({\em solid line}). The vertical line
shows the J08 distance limit. The overdensities in W1 ({\em top left}) and W2
({\em top right}) beams are due to the Sagittarius and Monoceros streams,
respectively. Beyond $R_{gal}\sim35$ kpc, the broken power law model provides a
much better fit to the data than the single power law J08 model (e.g., the J08
model overpredicts the halo stellar number density by $\gtrsim50\%$ in W2 and W3
beams at $R_{gal}\gtrsim35$ kpc).
\label{profiles.r.diff.SJI10}}
\end{figure}

Beyond $R_{gal} \sim 25$ kpc, in both W3 and W4 beams, the J08 model
overpredicts the observed counts. In particular, the J08 model overpredicts the
density observed in the final two bins of W3 beam by 20\% and 40\%,
respectively. Since we have measured incompleteness to be on the order of
$\sim5$\% in almost the entire observed range (see Figure~\ref{classification}),
and have made the bins volume-complete, this turnover cannot be a result of
observational bias but an indication of a change in halo density profile beyond
$R_{gal} \gtrsim 25$ kpc. The overprediction by the model may be even greater
since the observed CFHTLS stellar counts are {\em overestimated} by 15-20\% at
the faint end due to the contamination by galaxies, as shown in
Figure~\ref{classification}.

The profile exhibited by field W2 is also unusual. As shown in
Figure~\ref{profiles.r.J08}, it is significantly steeper than either W3 or W4
profile or the J08 prediction. By itself it is well described by a single
$n=-4.5, q=0.65$ power law. This fit, however, is incompatible with observations
from the other three directions where a shallower profile closer to $n\sim-2.8$ is strongly preferred. The curious behavior appears to be a combination of two
effects: i) the overdensity created by the Monoceros stream at
$R_{gal} \lesssim 25$ kpc and ii) the steepening of the halo profile beyond
$R_{gal} \sim 30$ kpc.

The Monoceros stream \citep{new02} is present in the general direction of W2
field ($l = 232\arcdeg$, $b=26\arcdeg$). For example, in an AAT/WFI survey of
the anticenter region, \citet{con07} detect the stream in the $l=220\arcdeg$,
$b=15\arcdeg$ direction at $D=11\pm1.6$ kpc, as well as at $l=240\arcdeg$,
$b=10\arcdeg$ with $D=13.8$ kpc. Similarly, J08 are able to trace the stream in
SDSS star count maps to $l\sim 230\arcdeg$ and importantly, show that it extends
to at least $b\sim 25\arcdeg$ of Galactic latitude where it forms a factor of
$\sim 1.5-2$ overdensity with respect to the extrapolation of the smooth stellar
background before and after the stream (e.g., as seen in the top left panel of
Figure 13 in J08). This is consistent with the factor of $\sim 2$ overdensity
observed in W2. Secondly, both of these studies estimate the Galactocentric
distance and width of the stream of $R_{gal} \sim 18$ kpc and
$\Delta R_{gal} \sim 3-4$ kpc in direction of W2. Given that these
characteristics are broadly consistent with the enhancement in the W2 direction
for $R_{gal} \lesssim 25$kpc, and having no additional evidence to the contrary,
we interpret the observed enhancement as a detection of the Monoceros stream.

\section{Detection of a Break in the Halo Density Profile\label{sec.break}}

Taking into account the enhancements due to the Sagittarius and Monoceros
streams, a single power law remains an appropriate description of the smooth
halo component  to Galactocentric distances of $R_{gal} \sim 25-30$ kpc. Beyond
this limit, however, the observed profile appears to turn over rather quickly
and the model needs to be modified to explain it.

To assess the character of the observed turnover, we fit a series of models of 
increasing complexity to the observed data. We use $\chi^2$ per degree of
freedom ($\chi^2_{dof}$) as the goodness of fit metric, and search for minima in
$\chi^2$ hypersurfaces using a Levenberg-Marquardt non-linear solver as
implemented by the GNU Scientific Library\footnote{GSL version 1.13;
\url{http://www.gnu.org/software/gsl/}}. To increase the likelihood of finding
the true global minimum, we repeat the minimization procedure with 10,000
different initial conditions selected randomly from a plausible range of initial
values of each parameter.

We begin by fitting a single J08-type power law to all admissible data
points\footnote{Those having $|Z|>5$kpc to avoid any contamination by the
disk.} of beams W3 and W4, the first eight data points of W1 beam (those that
show no contamination by the Sagittarius stream), and the last six data points
of W2 beam (those that we judge are past the influence of the Monoceros stream).
We obtain $\rho_0 = 1.7\times 10^{-6}$ pc$^{-3}$, $q=0.72$, $n=-2.9$ as the
best, but less satisfactory ($\chi^2_{dof} = 6.8$), fit. The $\rho_0$ is the
normalization (number of stars per pc$^{3}$) for the $0.2<g-r<0.3$ color bin we
use. For comparison, the fiducial model with J08 parameters shown in
Figure~\ref{profiles.r.J08} has $\chi^2_{dof} = 9.1$ when fitted to the same
data.

We next increase the complexity of the model by allowing for triaxiality of the
ellipsoid, parametrized by $w \equiv b/a$ (the ratio of ellipsoid axes):
\begin{equation}
 \rho(x, y, z) \propto (x^2 + \frac{y}{w}^2 + \frac{z}{q}^2)^{\frac{n}{2}}
\end{equation}
This addition makes practically no difference; the best-fit model changes only
slightly ($\rho_0 = 1.7\times 10^{-6}$ pc$^{-3}$, $q=0.72$, $n=-2.96$, $w=1.02$)
while $\chi^2_{dof}$ actually increases (to $\sim 7$) because of the extra
degree of freedom.

We continue by permitting the triaxial halo ellipsoid to rotate in the $X-Y$
(Galactic) plane by an angle $\phi$:
\begin{eqnarray}
x' & = & \cos(\phi)\, x - \sin(\phi)\, y \nonumber \\ 
y' & = & \sin(\phi)\, x + \cos(\phi)\, y \nonumber \\ 
 \rho(x, y, z) & \propto & (x'^2 + \frac{y'}{w}^2 + \frac{z}{q}^2)^{\frac{n}{2}} 
\end{eqnarray}
This five-parameter model marginally improves the fit
($\chi^2_{dof} = 6.8$), but converges to parameters $\rho_0 = 1\times 10^{-6}$
pc$^{-3}$, $q=1.11$, $n = -3.3$, $w=1.3$, $\phi=230\arcdeg$ that are strongly
excluded by prior data (e.g., \citealt{che01}, J08, and others).

Given the results of this series of experiments, a single power law is unlikely
to describe the observed counts: the data require a functional form allowing
for a change in the radial profile beyond $R_{gal} \gtrsim 25$ kpc. We therefore
attempt a series of simple ``broken power law'' models, where the density
follows one (the ``inner'') power law until radius $R_{\rm br}$ is reached, and
the other (the ``outer'' power law) beyond.

We begin with a minimal extension of a single power law model, allowing for a
change of the power law index beyond a certain radius $R_{\rm br}$:
\begin{eqnarray}
          R_e &       = & (x^2 + y^2 + \frac{z}{q}^2)^{\frac{1}{2}} \nonumber \\
\rho(x, y, z) & \propto & \left\{
			    \begin{array}{c}
			    (R_e)^{n_{\rm inner}}, \;\;\;\;\;\;\; R_e < R_{\rm br} \\
			    (R_e)^{n_{\rm outer}}, \;\;\;\;\;\;\; R_e > R_{\rm br} \\
\end{array}
\right.
\end{eqnarray}
Note that, as the ellipsoid is allowed to be oblate or prolate, the break radius
$R_{\rm br}$ is only equal to the physical Galactocentric radius $R_{gal}$ on
the $x$ axis (along the line connecting the Galactic center and the Sun). In the
vertical direction, the physical radius corresponding to $R_{\rm br}$ is reduced
by a factor of $q^{-1}$.

The above model, with five free parameters, produces a significantly better fit
to the data ($\chi^2_{dof} = 3.9$). The best-fit parameters for the inner power
law, $\rho_0 = (1.45\pm0.05)\times 10^{-6}$ pc$^{-3}$,
$q=0.70\pm0.01$, $n_{\rm inner} = -2.62\pm0.04$, $R_{\rm br} = 27.8\pm0.8$ kpc,
$n_{\rm outer} = -3.8\pm0.1$ are in excellent agreement with the J08 model,
while beyond $R_{\rm br}=27.8$ kpc the best-fit profile becomes steeper than the
J08 model.

As shown in Figure~\ref{profiles.r.diff.SJI10}, the observed profiles are
better fit by the broken power law than by the J08 model, excluding regions with
known tidal streams. A fit that entirely excludes W1 and W2 fields (not just the
regions with known tidal streams) does not strongly constrain the broken power
law model. In this fit, the normalization $\rho_0$ has high ($\sim50\%$)
fractional uncertainty and strongly correlates with oblateness $q$ and break
radius $R_{\rm br}$ (correlation coefficient is $\sim1$). This strong
correlation is caused by similar positions of W3 and W4 beams in $x-z$ and $y-z$
planes with respect to the Galactic plane (see Figure~\ref{xsections}). In
comparison, our best-fit profile that uses all four beams while excluding
regions with known tidal streams shows much weaker correlation between $\rho_0$,
$q$, and $R_{\rm br}$ ($\sim0.4$).

The inner parts of the W3 field in our best-fit model do show a systematic
underestimate of the counts by the model on the order of $\sim 15$\% (at an
approximately $1\sigma$-$2\sigma$ level\footnote{Note, however, that the error
bars of adjacent bins are highly correlated by observational errors.}), while
the opposite occurs in inner parts of the W1 field. These lines of sight point
toward high latitudes in the northern and southern Galactic hemispheres, and
the observed difference may be a signature of slight north-south asymmetry of
halo star counts recently seen in SDSS III data (Bonaca, A. et al.(2011), in
preparation).

To further assess the robustness of the detected break, we examined two more
variants of the model: the first, where we allowed the outer halo to have an
oblateness parameter $q$ different from that of the inner halo, and the second,
where we fixed the parameters of the inner halo to J08 values, and allowed only
those of the outer halo to vary. Both cases resulted in a similar value of
$\chi^2_{dof}$, as well as similar break radii ($R_{\rm br} \sim 28$ kpc) and
outer power law indices ($n_{\rm outer} \sim -3.8$). Importantly, the model with
varying $q_{\rm inner}$ and $q_{\rm outer}$ produced best-fit values of $0.71$
and $0.69$ for the two, respectively, indicating that there is no evidence for a
change in oblateness of the halo across the range of distances examined.

\begin{figure}
\epsscale{1.0}
\plottwo{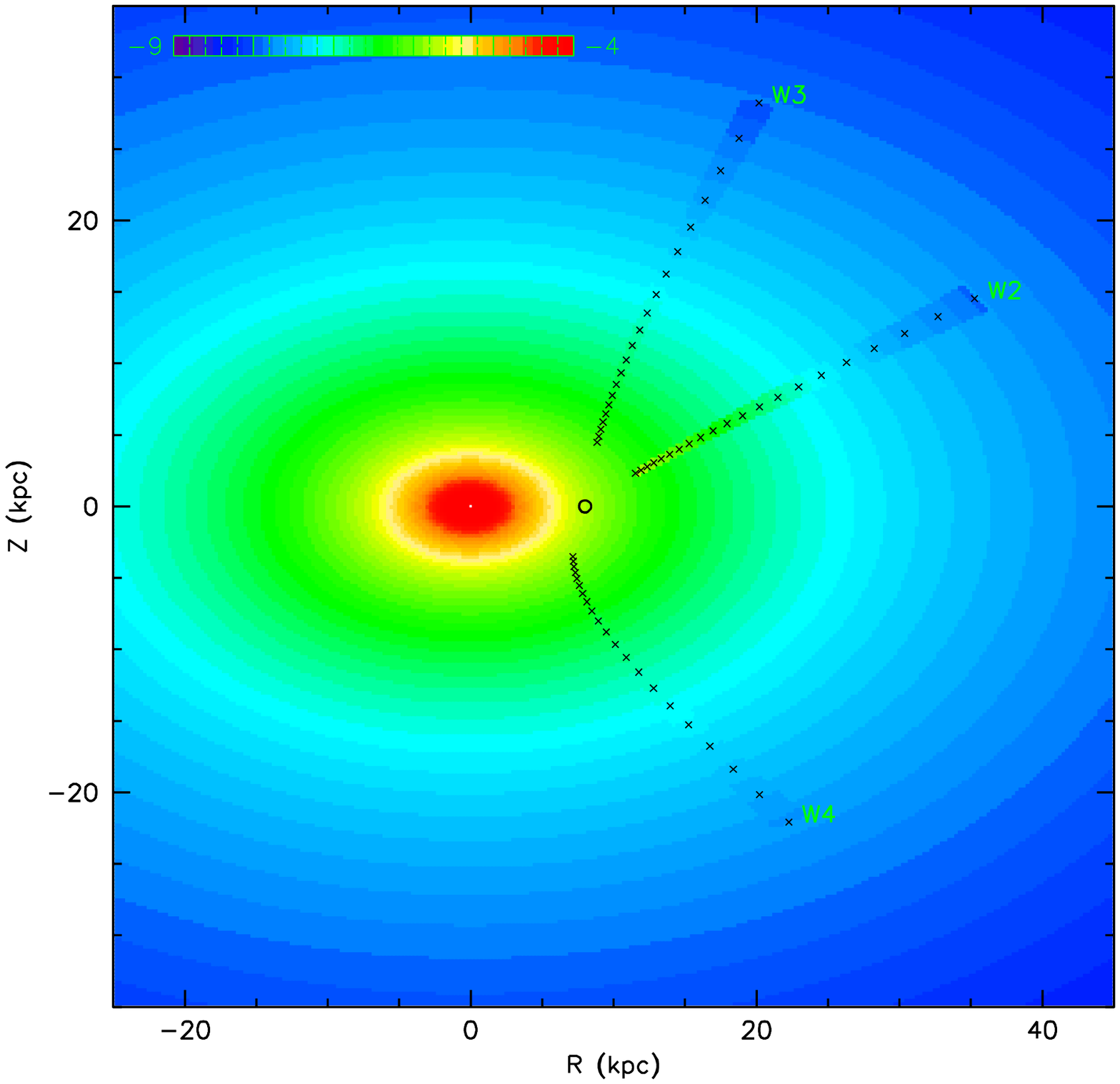}{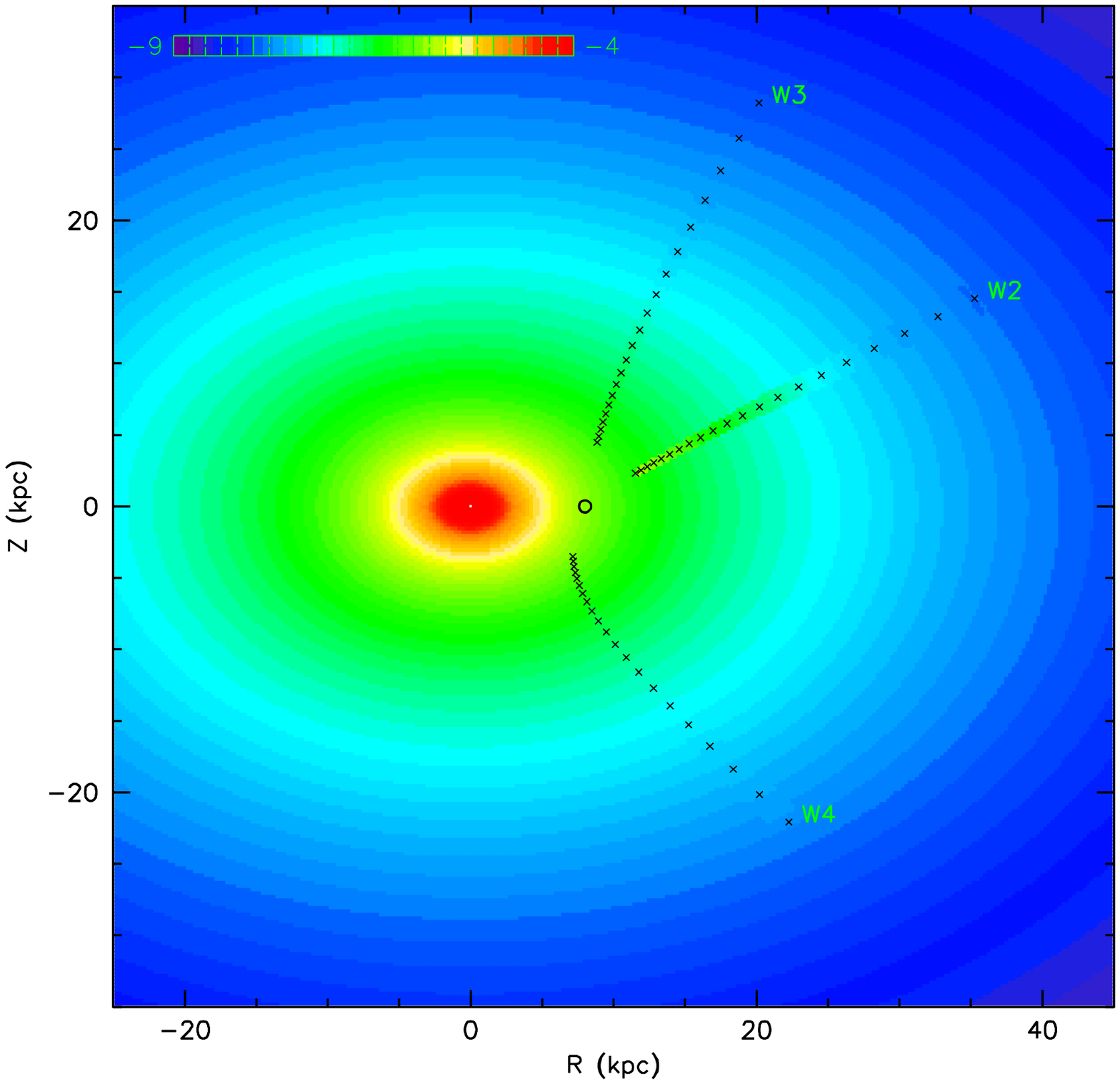}
\caption{
$R-Z$ plane visualization of the J08 power law halo model (left) and the broken
power law model presented in this paper (right). The color encodes the logarithm
of the number density of halo stars (stars pc$^{-3}$) predicted by the model.
Overplotted are the densities derived from the analysis of CFHTLS data (beams
W2, W3 and W4) presented in this paper (W1 beam is not shown because of the
strong contamination by the Sagittarius stream). Note the marked improvement in
data-model agreement for the broken power law model.
\label{rzplots}}
\end{figure}

Based on the above analysis, we conclude that the detection of steepening of the
halo density power law is robust. To explain the CFHTLS data, the power law
index needs to change from $n_{\rm inner} = -2.62$ to $n_{\rm outer} = -3.8$
around $R_{\rm br} \sim 28$ kpc. An R$-$Z plane visualization of the J08
power-law halo model and the broken power-law model presented in this paper is
shown in Figure~\ref{rzplots}.

We also fit Einasto's model \citep{ein65} to our data to allow easy comparison
with density profiles obtained from $N$-body simulations
\citep{nav04,die04,mer05,gra06}. The best-fit parameters for the Einasto's model
are $n=2.2\pm0.2$, $R_e=22.2\pm0.4$ kpc, $\rho_0=1.06\pm0.05$ stars pc$^{-3}$,
and $q=0.70\pm0.01$ with $chi^2_{pdf}=4.25$.

Due to contamination by galaxies at the faint end (see
Figure~\ref{classification}), the $n_{\rm outer} = -3.8\pm0.1$ power law index
given above is likely somewhat shallower than what it should be. To estimate by
how much $n_{\rm outer}$ is shallower due to contamination by galaxies, we
determine $f(r)= [1-contamination(r)]/completeness(r)$ for each star, where $r$
is a star's $r$ band magnitude, and $completeness(r)$ and $contamination(r)$
are the solid and dashed lines from Figure~\ref{classification}, respectively.
We sum $f(r)$ values in distance modulus bins and use them instead of raw number
counts when fitting the model. With this approach, a star located in a distance
modulus bin with higher galaxy contamination will contribute a value smaller
than one towards the total count in that bin. Finally, we find that
$n_{\rm outer}$ decreases by 0.1 to $n_{\rm outer}=-3.9$, which is still within
the uncertainty of $n_{\rm outer}=-3.8\pm0.1$ determined previously.

\section{Metallicity Distribution in the Halo and its Impact on the Best-fit Model\label{metallicity}}

In addition to stellar number density distribution, the metallicity distribution
can also provide strong constraints for halo formation models. Thanks to CFHTLS
$u$ band data, it is possible to compute photometric metallicity using a method
developed by \citet{ivezic08a}. Furthermore, the absolute magnitude of a star,
and hence its distance, depend on the star's metallicity (Equation~\ref{M_r})
and thus it is important to study systematic errors in best-fit model parameters
as a function of metallicity distribution.

\begin{figure}
\epsscale{1.0}
\plotone{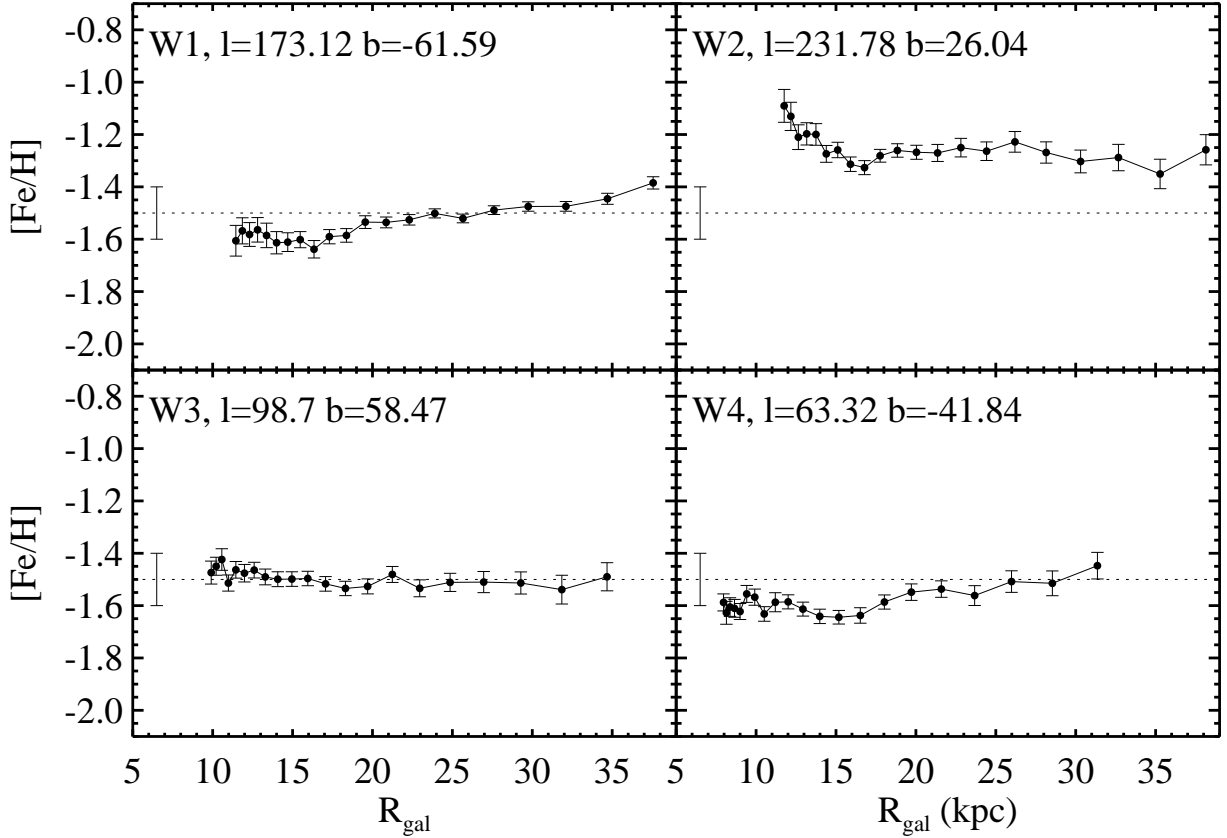}
\caption{
Median photometric metallicity ({\em symbols with error bars}) measured in
four CFHTLS wide survey beams as a function of distance from the Galactic
center, $R_{gal}$. The error bars show error in the median and the error bar at
(6.5, -1.5) shows the systematic uncertainty in the adopted photometric
metallicity method ($\sim0.1$ dex, \citealt{ivezic08a}). Within $R_{gal}\sim30$
kpc, the median metallicity is independent of distance and ranges from
$-1.4<[Fe/H]<-1.6$. The change in metallicity at $R_{gal}\sim15$ kpc, reported
by \citet{car07} and \citet{dej10}, is not evident. Apparently higher
metallicity in the W2 beam ($[Fe/H]\sim-1.3$ dex) may be due to $u$ band
calibration issues (see the text for a discussion).
\label{FeH_D}}
\end{figure}

Figure~\ref{FeH_D} shows the median halo metallicity in four CFHTLS beams as a
function of distance from the Galactic center, $R_{gal}$. In the two beams where
significant substructure is not detected (W3 and W4 beams), the median
metallicity is independent of distance and averages to $[Fe/H]\sim-1.5$ dex with
a range of 0.1 dex. This average value is consistent with the median halo
metallicity measured by \citet{ivezic08a} using SDSS F- and G-type main sequence
stars, and the range is consistent with the systematic uncertainty inherent to
this photometric metallicity method ($\sim0.1$ dex, \citealt{ivezic08a}).

The average metallicity in the W1 beam is also independent of distance and
averages $[Fe/H]\sim-1.5$ dex. This trend is probably a coincidence since the
Sagittarius tidal stream passes through the beam and the stream's metallicity is
not required to be $[Fe/H]\sim-1.5$ dex everywhere. On the contrary, models and
observations (\citealt{cho07}; \citealt{sesar10}; \citealt{lm10} and references
therein) suggest that the metallicity exhibits a gradient along the Sagittarius
tidal stream --- we simply happen to observe the stream where its metallicity
is $[Fe/H]\sim-1.5$ dex.

The metallicity in the W2 beam is a bit higher ($[Fe/H]\sim-1.3$ dex) than in
other beams, even at large Galactocentric distances ($R_{gal}>28$ kpc) where,
according to Figure~\ref{profiles.r.diff.SJI10} (top right panel), the
contribution of the Monoceros stream stars should be small. Since the
\citet{ivezic08a} photometric metallicity method depends on accurate $u$ band
photometry, anything affecting $u$ band measurements will also affect the
photometric metallicity estimate. As discussed in
Section~\ref{photometry_check}, the $u$ band measurements synthesized from
CFHTLS data may be impacted by some calibration issues in this beam, so these
issues are likely responsible for the apparently higher metallicity in the W2
beam (a $+0.05$ mag systematic offset in the $u$ band will increase the
metallicity by 0.2 dex).

Our observation that the halo metallicity is independent of Galactocentric
distance goes contrary to \citet{car07} and \citet{dej10} (hereafter deJ10)
results who conclude that the halo metallicity changes from $[Fe/H]\sim-1.6$ dex
to $[Fe/H]\sim-2.2$ dex at $R_{gal}\sim15$ kpc. To see how our best-fit model
varies when the metallicity changes from $[Fe/H]\sim-1.6$ dex for
$R_{gal}\lesssim15$ kpc to $[Fe/H]\sim-2.2$ dex beyond $R_{gal}\sim15$ kpc, we
iteratively modify the metallicity distribution in CFHTLS beams to reflect the
metallicity distributions shown in deJ10 Figure 7 (right, hereafter
$[Fe/H]_{deJ10}(D)$). New metallicities are assigned to CFHT stars by
interpolating $[Fe/H]$ from $[Fe/H]_{deJ10}(D)$ using initial heliocentric
distances, $D$ (calculated from Equations~\ref{M_r} and~\ref{photo_FeH}).
Heliocentric distances are then recalculated using new metallicity values, and
the metallicities are again interpolated from $[Fe/H]_{deJ10}(D)$). This process
is repeated (about 2-3 times) until the fractional difference between distances
in consecutive steps dips below 0.15 (fractional distance uncertainty for main
sequence stars is $\sim15\%$; \citealt{sij08}). The broken power law model is
fit to modified data once the convergence is achieved.

The best-fit model parameters obtained for deJ10-like metallicity distribution
are listed in Table~\ref{model_parameters}. In addition,
Table~\ref{model_parameters} also lists best-fit model parameters obtained for
constant metallicity distributions (i.e., all stars are assumed to have fixed
metallicity) and for the best-fit model presented in Section~\ref{sec.break}.
Within a range of plausible metallicity distributions, the best-fit model
parameters do not seem to vary greatly and average around $q\sim0.7$,
$n_{inner}\sim-2.4$, $n_{outer}\sim-3.8$, and $R_{br}\sim25$ kpc.

\section{Conclusions and Discussion\label{discussion}}

We have recalibrated CFHTLS ``wide'' survey
$u^*g^\prime r^\prime i^\prime i^\prime_2$ observations and transformed them to
the SDSS $ugri$ photometric system. Using a series of tests, we demonstrated
that synthesized $ugri$ observations, when compared to SDSS observations, show
no dependence on color or magnitude. The only exception to this are the $u$ band
observations synthesized from CFHTLS W2 field $u^*$ band observations, which
show a slight linear dependence on the $u-g$ color. Median photometric error
in synthesized $ugri$ photometry is $\sim0.03$ mag at the bright end and
$\sim0.1$ mag at $\sim22.5$ mag.

By obtaining the synthesized $ugri$ photometry from CFHTLS observations, we were
able to use the photometric parallax and metallicity relations which allowed us
to study the spatial and metallicity distribution of near-turnoff main sequence 
stars in the Galactic halo to heliocentric distances of $\sim35$ kpc. We find
that the halo number density profile becomes steeper at Galactocentric distances
greater than $R_{gal}\sim28$ kpc, with the power law index changing from
$n_{inner}=-2.62\pm0.04$ to $n_{outer}=-3.8\pm0.1$. While the best-fit model
parameters do change slightly depending on the adopted metallicity distribution
(see Table~\ref{model_parameters}), we find that a broken power law model is
required for a good fit to the data. We also find that the best-fit $R_{br}$
value cannot be smaller than $R_{br}=22$ kpc even for the most extreme
assumptions about the halo metallicity distribution (i.e., constant metallicity
at $[Fe/H]=-2.0$ dex from 5 to 35 kpc).

This study provides further evidence for the steepening of the halo density
profile previously detected by \citet{sesar10} using main sequence and RR Lyrae
stars from the SDSS stripe 82, and by \citet{keller08} using RR Lyrae stars from
the SEKBO survey. This result is consistent with predictions of galaxy formation
simulations which find a steepening of the density profile beyond $\sim30$ kpc
\citep{bj05,dh08,zol09}.

We see no evidence of change in halo metallicity within the range of probed
distances. The halo metallicity ranges between $[Fe/H]\sim-1.4$ dex and
$[Fe/H]\sim-1.6$ dex, and averages at $[Fe/H]\sim-1.5$ dex. This result runs
contrary to \citet{car07} and \citet{dej10} studies which report a metallicity
of $[Fe/H]\sim-1.6$ dex within $R_{gal}\lesssim15$ kpc, and $[Fe/H]\sim-2.2$ dex
beyond. Only the in situ spectroscopic metallicities of distant main sequence
stars may provide a definitive answer to this discrepancy. With the sky density
of near-MSTO stars at high Galactic latitudes of about 100 stars deg$^2$, the
multi-object capability over a wide field of view would be well matched to such
a program.

While the total sky coverage of the four CFHTLS beams is slightly smaller than
that of SDSS stripe 82 (220 deg$^2$ versus 300 deg$^2$), the CFHTLS beams
provide a much stronger constraint on the oblateness ($Z$ to $R$ semi-major axis
ratio) of the stellar halo because they probe very different Galactic lines of
sight. We find the oblateness to be $q=0.70\pm0.01$ and see no evidence of
change across the range of probed distances ($5<R_{gal}/{\rm kpc}<35$). This
result is quite consistent with the oblateness of the dark matter halo,
$q_{DM}=0.72$, obtained by \citet{lm10} using the positions and kinematics of
Sagittarius stream stars. However, we find the minor axis of the stellar halo to
be aligned with the spin vector of the Milky Way, while \citet{lm10} find the
minor axis of the dark matter halo to be {\em perpendicular} to the spin vector
of the Milky Way.

\begin{figure}
\epsscale{0.55}
\plotone{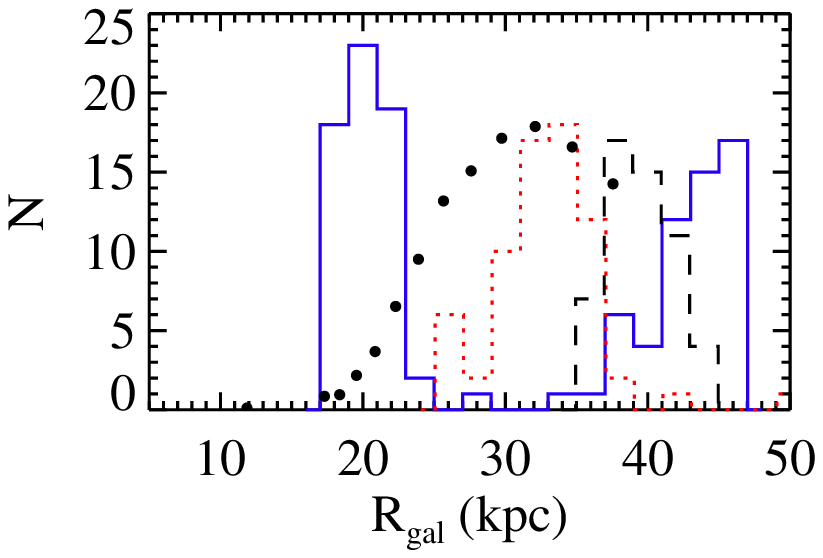}

\plotone{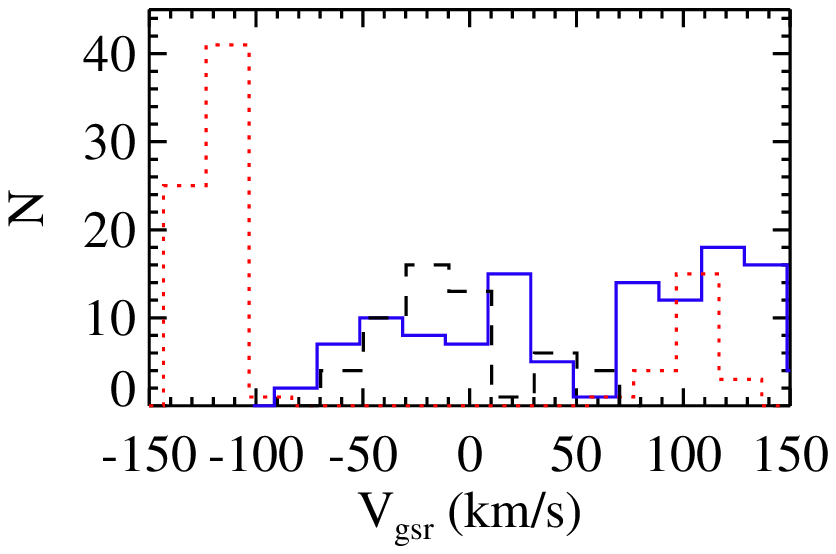}
\caption{
Galactocentric distance ({\em top}) and Galactocentric rest-frame radial
velocity distribution ({\em bottom}) of Sagittarius stream stars in the CFHTLS
W1 beam, as predicted by the \citet{lm10} model. The solid and dotted lines
denote stars in the first wrap-around leading and trailing tidal streams,
respectively. The dashed line represents stars in the leading arm that has
wrapped more than $360\arcdeg$ around the Milky Way from Sagittarius (second
wrap-around). The solid circles show the observed distribution of Sagittarius
tidal stream stars in the W1 beam obtained from ${\rm data}/{\rm model} - 1$
residuals of the broken power law model (multiplied by 10). The bottom panel
shows that the contribution of different Sagittarius streams to the observed
distribution at $R_{gal}\sim32$ kpc, overplotted in the top panel, can be more
easily quantified in the velocity space, since the majority of Sagittarius
trailing stream stars are predicted to narrowly distribute around $V_{gsr}<-130$ km s$^{-1}$, while the leading stream stars have a much broader velocity
distribution and $V_{gsr}>-100$ km s$^{-1}$.
\label{Sgr_hists}}
\end{figure}

We have detected the Sagittarius and Monoceros streams as excesses of stars in
CFHTLS W1 and W2 fields, respectively. These detections provide new constraints
on models of these streams. For example, the \citet{lm10} model of the
Sagittarius dwarf spheroidal galaxy predicts positions and velocities of
Sagittarius stream stars in the CFHTLS W1 beam (Figure~\ref{Sgr_hists}). The
spatial distribution of Sagittarius stars is predicted to be bimodal, with a
narrow peak at $R_{gal}\sim20$ kpc, and a broader peak at $R_{gal}\sim35$ kpc.
The observed distribution of Sagittarius stars in this region, overplotted in
Figure~\ref{Sgr_hists} (top panel), has only one peak at $R_{gal}\sim35$ kpc.
Unfortunately, we do not have radial velocity measurements in this region, and
cannot quantify contributions of particular streams (leading and trailing) to
this peak. Multi-epoch surveys, such as the Palomar Transient Factory (PTF;
\citealt{law09}) and LSST, will enable robust identification of bright tracers
such as RR Lyrae stars, which can be used to map the velocity structure in this
region. The resulting improvement in models of these streams may then help to
further constrain the shape, orientation, and the mass of the Galactic dark
matter halo.

\acknowledgments

B.~Sesar thanks NSF grant AST-0908139 to J.~G.~Cohen and NSF grant AST-1009987
to S.~R.~Kulkarni for partial support. \v{Z}.~Ivezi\'c acknowledges support by
NSF grants AST-0707901 and AST-1008784 to the University of Washington, by NSF
grant AST-0551161 to LSST for design and development activity, and by the
Croatian National Science Foundation grant O-1548-2009. M.~Juri\'c acknowledges
support by NASA through Hubble Fellowship grant HF-51255.01-A awarded by the
Space Telescope Science Institute, which is operated by the Association of
Universities for Research in Astronomy, Inc., for NASA, under contract NAS
5-26555. Partial support for this work was provided by NASA through a contract
issued by the Jet Propulsion Laboratory, California Institute of Technology
under a contract with NASA. Based on observations obtained with
MegaPrime/MegaCam, a joint project of CFHT and CEA/DAPNIA, at the
Canada-France-Hawaii Telescope (CFHT) which is operated by the National Research
Council (NRC) of Canada, the Institut National des Science de l'Univers of the
Centre National de la Recherche Scientifique (CNRS) of France, and the
University of Hawaii. This work is based in part on data products produced at
TERAPIX and the Canadian Astronomy Data Centre as part of the
Canada-France-Hawaii Telescope Legacy Survey, a collaborative project of NRC and
CNRS.

\clearpage


\begin{deluxetable}{cccccc}
\tabletypesize{\scriptsize}
\tablecolumns{6}
\tablewidth{0pc}
\tablecaption{Overview of the CFHTLS Wide fields\label{wide_fields}}
\tablehead{
\colhead{CFHTLS Field Name} & \colhead{$\alpha^a$ $(deg)$} &
\colhead{$\delta^a$ $(deg)$} & \colhead{l$^b$ $(deg)$} &
\colhead{b$^b$ $(deg)$} & \colhead{Sky Coverage $(deg^2)$} 
}
\startdata
W1 &  34.5 &   -7.0 & 173.12 & -61.59 & 72 \\
W2 & 134.5 & -3.3 & 231.78 &  26.04 & 25 \\
W3 & 214.4 & 54.5 & 98.70 &  58.47 & 49 \\
W4 & 333.3 &  1.3 & 63.32 & -41.84 & 25
\enddata
\tablenotetext{a}{Equatorial J2000.0 right ascension and declination.}
\tablenotetext{b}{Galactic longitude and latitude.}
\end{deluxetable}

\clearpage

\begin{deluxetable}{ccccccc}
\tabletypesize{\scriptsize}
\tablecolumns{7}
\tablewidth{0pc}
\tablecaption{$\Delta_{f}^{abs}$ Values for CFHTLS ``Wide'' Fields\label{abs_coeff}}
\tablehead{
\multicolumn{3}{c}{} & \multicolumn{4}{c}{$\Delta_{f}^{abs}$} \\
\cline{1-7} \\
\colhead{CFHT Band} & \colhead{SDSS Color Range} & \colhead{C} &
\colhead{W1} & \colhead{W2} & \colhead{W3} & \colhead{W4}
}
\startdata
$u^*$        & $1.1<u-g<1.4$ & -0.155 &  0.087 &  0.099 &  0.098 &  0.097 \\
$g^\prime$   & $0.3<g-r<1.2$ & -0.161 & -0.017 & -0.044 & -0.037 & -0.043 \\
$r^\prime$   & $0.4<g-r<1.2$ & -0.018 & -0.030 & -0.019 & -0.027 & -0.027 \\
$i^\prime$   & $0.3<r-i<1.0$ & -0.081 & -0.020 & -0.015 & -0.022 & -0.030 \\
$i^\prime_2$ & $0.3<r-i<1.0$ &  0.020 &      0 &    n/a &    n/a &   0    
\enddata
\end{deluxetable}

\clearpage

\begin{deluxetable}{ccccc}
\tabletypesize{\scriptsize}
\tablecolumns{5}
\tablewidth{0pc}
\tablecaption{$\Delta_{j,f}^{total}$ Values for the CFHTLS W1$-$W4 fields\label{W1_total_coeff}}
\tablehead{
\colhead{Pointing$^a$} & \colhead{$u^*$} & \colhead{$g^\prime$} &
\colhead{$r^\prime$} & \colhead{$i^\prime$}
}
\startdata
W1+0-2  & 0.106 & -0.034 & -0.034 & -0.048 \\
W1+0-3  & 0.105 & -0.030 & -0.028 & -0.040 \\
W1+0-4* & 0.077 & -0.023 & -0.026 & 0.001
\enddata
\tablenotetext{a}{Pointings with names ending with ``*'' have $i^\prime_2$ instead of $i^\prime$ observations.}
\tablecomments{This table is available in its entirety in machine-readable
form in the online journal. A portion is shown here for guidance regarding its
form and content.}
\end{deluxetable}

\clearpage

\begin{deluxetable}{ccccccc}
\tabletypesize{\scriptsize}
\tablecolumns{7}
\tablewidth{0pc}
\tablecaption{Best-fit Model Parameters and their Uncertainties\label{model_parameters}}
\tablehead{
\colhead{Adopted Metallicity} & \colhead{$\rho_0^a$ ($10^{-6}$ pc$^{-3}$)} & \colhead{q$^b$} & \colhead{$n_{inner}$} & \colhead{$R_{br}$ (kpc)} & \colhead{$n_{outer}$} & \colhead{$\chi^2_{dof}$$^c$}}
\startdata
Equation~\ref{photo_FeH}$^d$ & $1.45\pm0.05$ & $0.70\pm0.01$ & $-2.62\pm0.04$ & $27.8\pm0.8$ & $-3.8\pm0.1$ & 3.9 \\
\citet{dej10}$^e$            & $1.25\pm0.04$ & $0.76\pm0.01$ & $-2.32\pm0.04$ & $24.3\pm0.3$ & $-6.4\pm0.1$ & 9.6 \\
$[Fe/H]=-1.0^f$              & $1.02\pm0.04$ & $0.72\pm0.01$ & $-2.32\pm0.04$ & $28.5\pm0.6$ & $-3.8\pm0.1$ & 5.7 \\
$[Fe/H]=-1.5^f$              & $1.32\pm0.05$ & $0.73\pm0.01$ & $-2.47\pm0.05$ & $24.8\pm0.7$ & $-3.86\pm0.08$ & 4.4 \\
$[Fe/H]=-2.0^f$              & $1.51\pm0.06$ & $0.72\pm0.01$ & $-2.50\pm0.06$ & $22.1\pm0.6$ & $-3.81\pm0.06$ & 4.6
\enddata
\tablenotetext{a}{Number density of halo stars with $0.2 < g-r < 0.3$ at the position of the Sun.}
\tablenotetext{b}{Oblateness ($Z$ to $R$ semi-major axis ratio).}
\tablenotetext{c}{Goodness of the fit (smaller is better).}
\tablenotetext{d}{Our best-fit model. Metallicity calculated for each star using Equation~\ref{photo_FeH}.}
\tablenotetext{e}{Metallicity distribution adopted from \citet{dej10} Figure 7 (right).}
\tablenotetext{f}{Metallicity fixed for all stars at a given value.}
\end{deluxetable}

\clearpage

\clearpage

\clearpage

\clearpage

\clearpage

\clearpage

\clearpage

\clearpage

\clearpage

\clearpage

\clearpage

\clearpage

\clearpage

\end{document}